\documentclass[preprint,12pt,authoryear]{elsarticle}

\usepackage{amsmath,amssymb}
\usepackage[english]{babel}
\usepackage{natbib, url}
\usepackage{xcolor}
\usepackage{float,graphicx,subfigure}

\newcommand{\Fig}[1]{Fig. \ref{#1}}
\newcommand{\Sec}[1]{Sect. \ref{#1}}
\newcommand{\Eq}[1]{Eq.~(\ref{#1})}
\newcommand{\pref}[1]{(\ref{#1})}
\newcommand{\Tab}[1]{Tab. \ref{#1}}

\newcommand{\D}{\,\mathrm{d}}
\newcommand{\icp}{\,\mathrm{icp}}

\bibpunct{(}{)}{,}{}{}{,}
\journal{Journal of Mathematical Psychology}

\begin{document}

\begin{frontmatter}

\title{Quantum approaches to music cognition}
\author[btu]{Peter beim Graben\corref{cor1}}
\ead{peter.beimgraben@b-tu.de}
\author[ua]{Reinhard Blutner}
\cortext[cor1]{Corresponding author}

\address[btu]{Brandenburgische Technische Universit\"{a}t Cottbus -- Senftenberg, \\
    Institute of Electronics and Information Technology, \\
    Department of Communications Engineering, \\
    Cottbus, Germany}
\address[ua]{Emeritus Universiteit van Amsterdam, \\
    ILLC, \\
    Amsterdam, Netherlands}

\begin{abstract}
Quantum cognition emerged as an important discipline of mathematical psychology during the last two decades. Using abstract analogies between mental phenomena and the formal framework of physical quantum theory, quantum cognition demonstrated its ability to resolve several puzzles from cognitive psychology. Until now, quantum cognition essentially exploited ideas from projective (Hilbert space) geometry, such as quantum probability or quantum similarity. However, many powerful tools provided by physical quantum theory, e.g., symmetry groups have not been utilized in the field of quantum cognition research sofar. Inspired by seminal work by Guerino Mazzola on the symmetries of tonal music, our study aims at elucidating and reconciling static and dynamic tonal attraction phenomena in music psychology within the quantum cognition framework. Based on the fundamental principles of octave equivalence, fifth similarity and transposition symmetry of tonal music that are reflected by the structure of the circle of fifths, we develop different wave function descriptions over this underlying tonal space. We present quantum models for static and dynamic tonal attraction and compare them with traditional computational models in musicology. Our approach replicates and also improves predictions based on symbolic models of music perception.
\end{abstract}

\begin{keyword}
music psychology, tonal attraction, quantum cognition, tonal space
\end{keyword}

\end{frontmatter}

\paragraph{Highlights}
\begin{itemize}
  \item Quantum models for tonal attraction are compared with symbolic models in musicology
  \item Symbolic models of tonal attraction are parsimoniously derived from fundamental symmetries of music cognition
  \item Tones appear as Gestalts, i.e. wave fields over the circle of fifths
\end{itemize}

\section{Introduction}
\label{sec:intro}

Tonal attraction is an important concept in music cognition. It refers to the idea that melodic or voice-leading pitches tend toward other pitches in greater or lesser degrees, and has been empirically investigated by means of \emph{probe tone experiments} \citep{KrumhanslShepard79, Krumhansl79}. This paradigm can be regarded as a priming experiment in music psychology: In each trial, a subject is presented with a priming context (e.g. a scale, a chord, or a cadence) that establishes a tonal key, say, C major, which is followed by a probe tone randomly chosen from the twelve tones of the chromatic scale (see \Sec{sec:musico} for details). Subjects are then asked to rate the amount of attraction exerted by the priming context upon the probe tone. Depending on the instruction, one can distinguish between probe tone experiments on static or dynamic attraction \citep{Parncutt11}. In the static attraction paradigm subjects are asked to rate \emph{how well the probe tone fits into} the preceding context \citep{KrumhanslKessler82}, while in the dynamic attraction paradigm subjects are asked to rate \emph{how well the probe tone completes or resolves} the preceding context \citep{Temperley08, Woolhouse09}.

The difference between these different instructions is illustrated as follows: In the static attraction experiment a C major context primes the C major key such that C (the \emph{tonic}) as a probe tone receives maximal attraction, followed by the members of the tonic triad G (the \emph{dominant}) and E (the \emph{mediant}). By contrast, in the dynamic attraction experiment a C major context primes the F major key because the chord CEG is the dominant triad of F major which is resolved by its tonic triad FAC, making the probe tone F most likely, followed by C and then A \citep{Woolhouse09}. Yet another possibility could be melodic progression where C primes its minor and major second intervals below and above, i.e. B and D within the C major scale according to the principle of \emph{pitch proximity} \citep{Temperley08}.

There are essentially two research lines for computational models of tonal attraction data. The first, psychoacoustic bottom-up models are inspired by Hermann von Helmholtz' \citeyearpar{Helmholtz1877} attempts on spectral representations and the overtone series \citep{MilneEA11, MilneLaneySharp15, Stolzenburg15}. The second, cognitive top-down models go back to Lerdahl and Jackendoff's \citeyearpar{LerdahlJackendoff83} generative theory of tonal music, using either hierarchically structured representations of tonal space \citep{Lerdahl88, Lerdahl96, KrumhanslKessler82, KrumhanslCuddy10} or statistical correlations in large music corpora that are modeled by probabilistic approaches, such as Gaussian Markov chains \citep{Temperley07, Temperley08}.

This study aims at integrating structural and probabilistic theories of computational music theory into a unified framework. On the one hand, structural accounts such as the generative theory of tonal music are guided by principle musicological insights about the intrinsic symmetries of Western tonal music \citep{Balzano80, LerdahlJackendoff83, Lerdahl88, Lerdahl96}. On the other hand, probabilistic accounts such as Bayesian models or Gaussian Markov chains are able to describe melodic progression through statistical correlations in large music corpora \citep{Temperley07, Temperley08}. Here, we apply the framework of quantum cognition \citep{PothosBusemeyer13, BusemeyerBruza12} to music cognition in order to unify symmetry and (quantum) probability in a single framework for data from static and dynamic attraction experiments.

The paper is structured as follows: In the next section \ref{sec:matmet} we present, after a brief recapitulation of the essential concepts of music theory, two tonal attraction experiments of \citet{KrumhanslKessler82} on static attraction and of \citet{Woolhouse09} on dynamic attraction. Section \ref{sec:theo} reports our theory for modeling these results. For the static attraction data we first review the classical hierarchical models of tonal space \citep{Lerdahl88, Lerdahl96, KrumhanslKessler82, KrumhanslCuddy10} (for other models, cf. \citet{Temperley07, Temperley08, Stolzenburg15}). Second, we present two quantum models based on the essential symmetries of the chromatic octave. Also for the dynamic attraction data, we first review one particular model of tonal space, the interval cycle model of \citet{Woolhouse09} and \citet{WoolhouseCross10}. Guided by this approach we subsequently develop our quantum model for dynamical attraction analogous to the static case. The following section \ref{sec:res} presents the results of our quantum models and of the canonical models for comparison. The paper closes with a discussion and a short conclusion of our main findings.

\section{Material and methods}
\label{sec:matmet}

As this study is devoted to symmetries in quantum models of static and dynamic tonal attraction, we first give a brief introduction into mathematical musicology \citep{Balzano80, Mazzola90, Mazzola02, MazzolaMannonePang16}.

\subsection{Elements of music}
\label{sec:musico}

It is well known since Hermann von Helmholtz' \citeyearpar{Helmholtz1877} groundbreaking studies that the human auditory system cannot distinguish between acoustic stimuli with pitch frequencies $f_0, f_1$ when their frequency ratios are separated by one octave: $f_1 / f_0 = 2$. This fundamental \emph{principle of octave equivalence} induces an equivalence relation of acoustic stimuli into \emph{pitch classes}, or \emph{chroma} \citep{Parncutt11} which circularly wind up throughout different octaves. Western and also Chinese music divides this continuous pitch class circle into twelve distinguished tones, or \emph{degrees} \citep{MazzolaMannonePang16}, comprising the chromatic scale shown in \Tab{tab:chromsc}.

\begin{table}[H]
  \centering
  \begin{tabular}{l*{13}{c}}
  \hline
  interval $j$ & 0 & 1 & 2 & 3 & 4 & 5 & 6 & 7 & 8 & 9 & 10 & 11 & 12 \\
  tone & C & C$\sharp$ & D & E$\flat$ & E & F & F$\sharp$ & G & A$\flat$ & A & B$\flat$ & B & C' \\
  \hline
\end{tabular}
  \caption{\label{tab:chromsc} Pitch classes as chromatic scale degrees.}
\end{table}

While pitch frequencies $f_0, f_1$ appear equivalent, when they are separated by one octave, i.e. $f_1 = 2 f_0$ is the first overtone of the fundamental frequency $f_0$, the second overtone $f_2 = 3 f_0$ and the first one $f_1$ yet sound similar as they are separated by a perfect fifth $f_2 / f_1 = 3/2$. This auditory \emph{principle of fifth similarity} lead to the historic development of the \emph{diatonic scales} of Western tonal music \citep{Helmholtz1877, Schonberg78}. In diatonic scales seven tones are arranged in a particular order of full-tone (major second) and semitone (minor second) steps. For the key of C major, these are the seven tones in ascending order C, D, E, F, G, A, B, C', namely the white keys of a piano keyboard with C' closing the octave. The tonic C is regarded as the most stable note. The same tones arranged after cyclic permutation, A, B, C, D, E, F, G, A', entail the \emph{relative minor} scale of (natural) A minor with tonic A.

\begin{figure}[H]
\centering
 \subfigure[]{\includegraphics[scale=0.45]{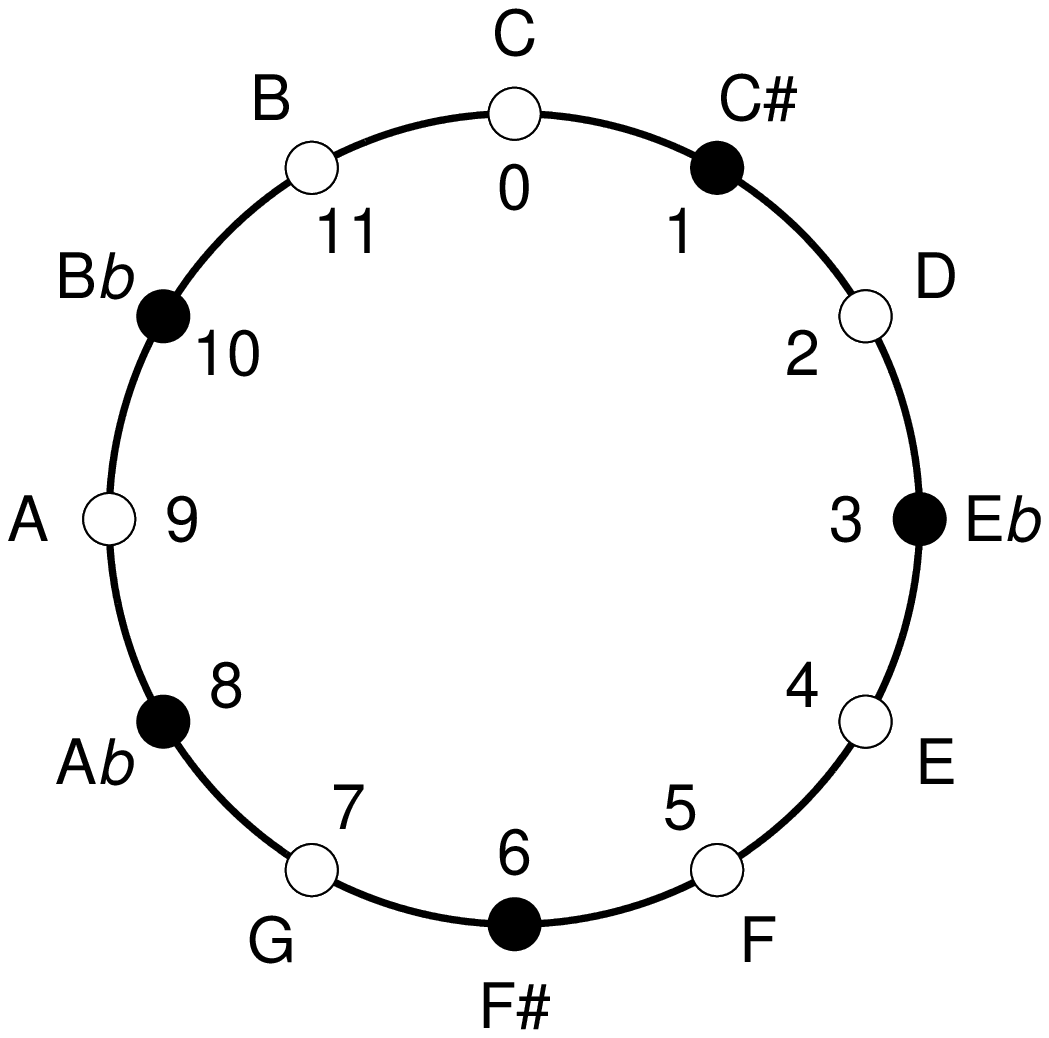}} \qquad
 \subfigure[]{\includegraphics[scale=0.45]{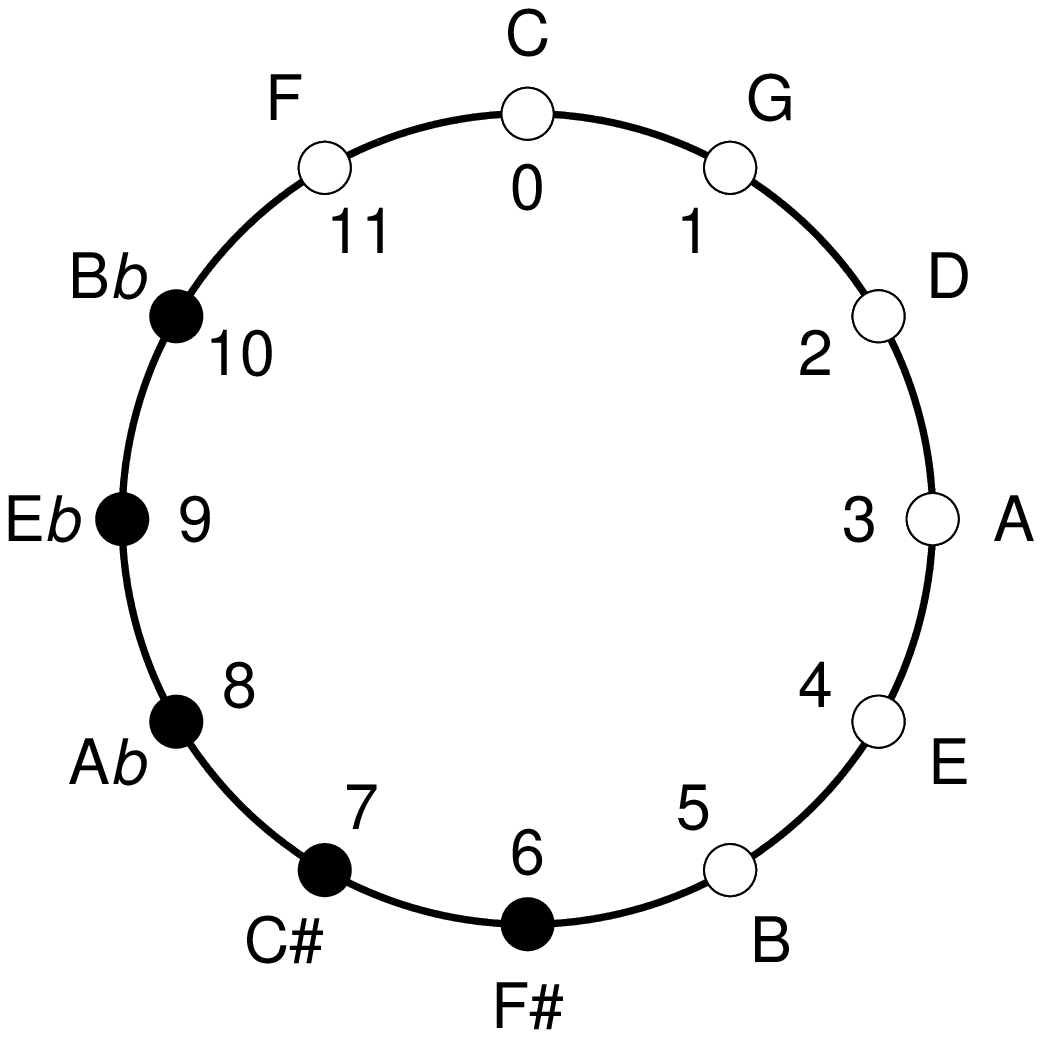}}
\caption{\label{fig:quinten} Tonal spaces of Western music. (a) Chroma circle of C major scale. (b) Circle of fifths.  Open bullets indicate scale (diatonic) tones; closed bullets denote non-scale (chromatic) tones.}
\end{figure}

Figure \ref{fig:quinten}(a) presents the chroma circle \citep{Krumhansl79} consisting of the twelve circularly repeating pitch classes of the chromatic scale in equal temperament, together with the C major scale degrees depicted as open bullets. The tonic C is indicated by $j = 0$. The relative A minor scale is obtained by rotating all tones 3 semitone steps (i.e. a minor third) in clockwise direction, thus assigning A to $j = 0$. The particular order: 2, 2, 1, 2, 2, 2, 1 of major second (2 steps) and minor second (1 step) intervals characterizes all major scale modes, while (natural) A minor is characterized by the interval order 2, 1, 2, 2, 1, 2, 2, as every other minor scale mode, too. Hence, the \emph{parallel minor} scale of C major is (natural) C minor, comprising C, D, E$\flat$, F, G, A$\flat$, B$\flat$, C' with tonic C.

The chroma circle \Fig{fig:quinten}(a) exhibits the geometric symmetry of a dodecagon, corresponding to the cyclic group of integer cosets modulo twelve, $\mathbb{Z}_{12}$ \citep{Balzano80, MazzolaMannonePang16}, that is generated by the semitone step $g \in \mathbb{Z}_{12}$ through its powers $g, g^2, g^3, \dots, g^{11}, g^{12}$ with $g^{12} = e$ as the neutral element of the group. Hence, the inverse of an element $g^k$ is given as $g^{12 - k}$ with $g^6$ as its own inverse, i.e. $g^6$ is nilpotent. The application of the generator $g \in \mathbb{Z}_{12}$ to the chroma circle corresponds to a clockwise transposition by one semitone in \Fig{fig:quinten}(a).

The cyclic group $\mathbb{Z}_{12}$ does not only possess the fundamental generator $g$. Yet all (integer) powers of $g$ that are relatively prime with the group order $n = 12$ are generators as well. Therefore, we obtain all generators of $\mathbb{Z}_{12}$ as $G_{12} = \{ g, g^{5},  g^{7}, g^{11} \}$. As $g$ corresponds to one semitone step in clockwise direction around the chroma circle, its inverse $g^{11} = g^{12 - 1}$ denotes one semitone step in counterclockwise direction. Similarly, $g^7$ in clockwise direction is inverse to $g^5$ in counterclockwise direction. Iterating the generator $g^7$ in clockwise direction to \Fig{fig:quinten}(a), i.e. by applying powers of $(g^7)^{m}$ to the tonic, yields the important \emph{circle of fifths}, shown in \Fig{fig:quinten}(b), where all diatonic degrees are accumulated in a connected set \citep{Balzano80}. With the tonic of C major in position $j=0$, its \emph{dominant} G is in position $j=1$, and its \emph{subdominant} F is in position $j=11$. The circle of fifths [\Fig{fig:quinten}(b)] reflects both fundamental principles of auditory pitch perception: octave equivalence is indicated by the identity $g^{12} = e$, whereas fifth similarity becomes geometrical neighborhood: two tones $(g^7)^{m}, (g^7)^{n}$ are the more similar the smaller $n - m \mod 12$. Alternatively, one could say that two tones are the more similar the smaller their angle along the circle of fifths. This similarity measure leads directly to our quantum model discussed in \Sec{sec:free}.

The universal symmetry in tonal music is \emph{transposition invariance} \citep{Kohler69}: A melody does not significantly alter its character, when played in a different key. Transposition from one key into another one is carried out by rotations either along the chroma circle or along the circle of fifths. Rotating here all tones one step in clockwise direction, assigns $j=0$ to F which thereby becomes the tonic of F major with dominant $j=1$ at C and subdominant $j=11$ at B$\flat$. Correspondingly, G major is obtained by a counterclockwise rotation by one step with tonic G ($j=0$), dominant D ($j=1$), and subdominant C ($j=11$).

These are only a few examples for symmetries of the chroma circle. Others are the dichotomy between consonances and dissonances \citep{Helmholtz1877, MazzolaMannonePang16}, or the emergence of the \emph{third torus} as the direct group product $\mathbb{Z}_{12} = \mathbb{Z}_{3} \times \mathbb{Z}_{4}$ \citep{Balzano80, MazzolaMannonePang16} which is crucial for the canonical construction of tertian chords in harmony theory \citep{Schonberg78}. This decomposition maps both, the chroma circle and the circle of fifths onto a torus, generated by a minor third (3 steps) in one direction and a major third (4 steps) in the orthogonal direction. Starting with C and going 4 steps, yields E, from where one arrives at G after 3 orthogonal steps, thus forming the C major triad CEG. Iterating first 3 and then 4 steps, instead, entails the C minor triad CE$\flat$G. Likewise, the F major triad is comprised by its tonic F, its major third, A, and the minor third C of A. For G major, one obtains correspondingly GBD.

\subsection{Static tonal attraction}
\label{sec:mmstatic}

In a celebrated probe tone experiment on static tonal attraction, \citet{KrumhanslKessler82} asked musically experienced listeners to rate how well, on a seven point scale, each note of the chromatic octave fitted with a preceding context, which consisted of short musical sequences, such as ascending scales, chords, or cadences, in major or minor keys. All stimuli were prepared as artificial Shepard tones \citep{KrumhanslKessler82, Woolhouse09, MilneLaneySharp15}, i.e. superpositions of pure sinusoids over five octaves with Gaussian amplitude envelopes.

\begin{figure}[H]
\centering
\subfigure[]{\includegraphics[scale=0.3]{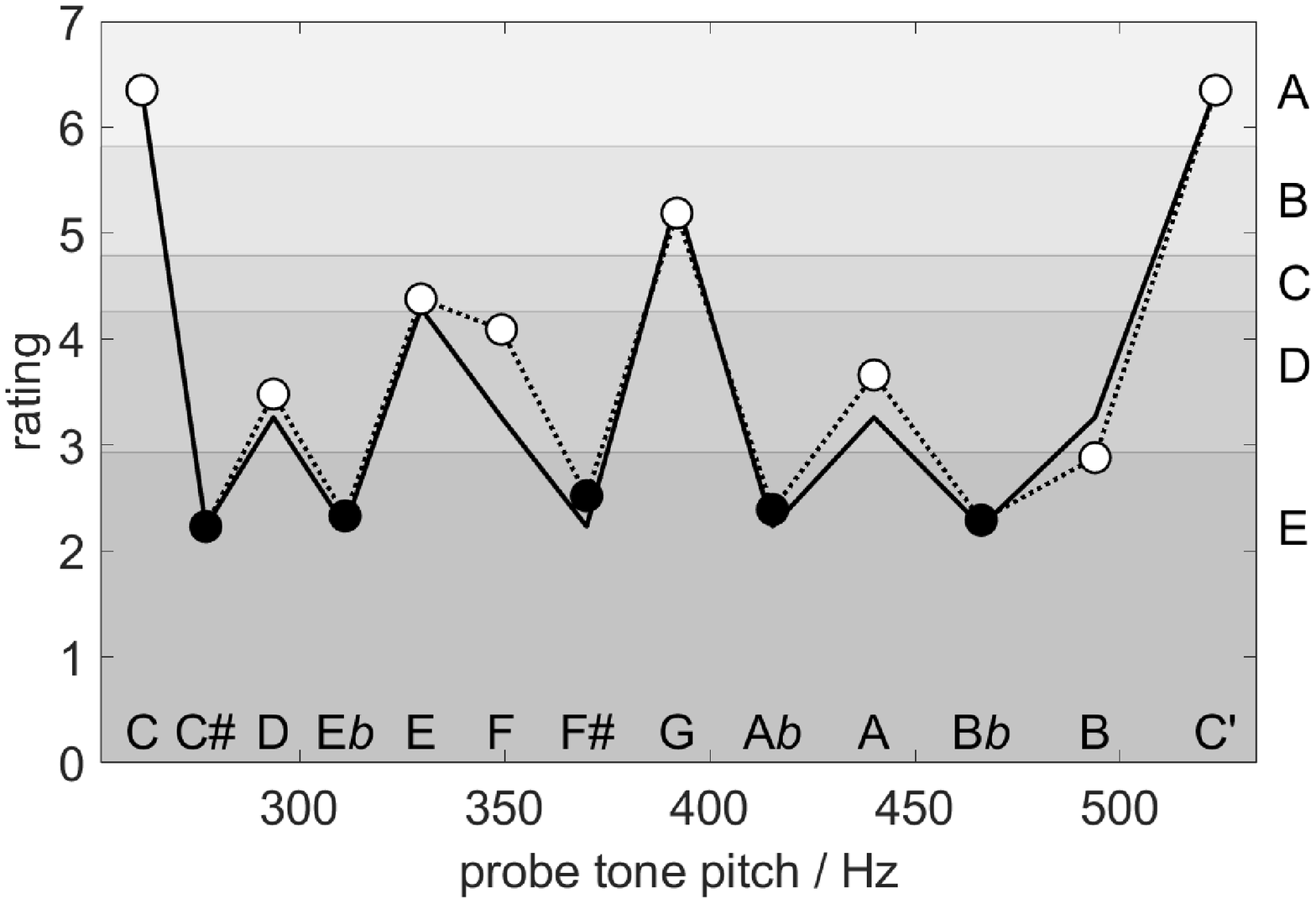}} \qquad
\subfigure[]{\includegraphics[scale=0.3]{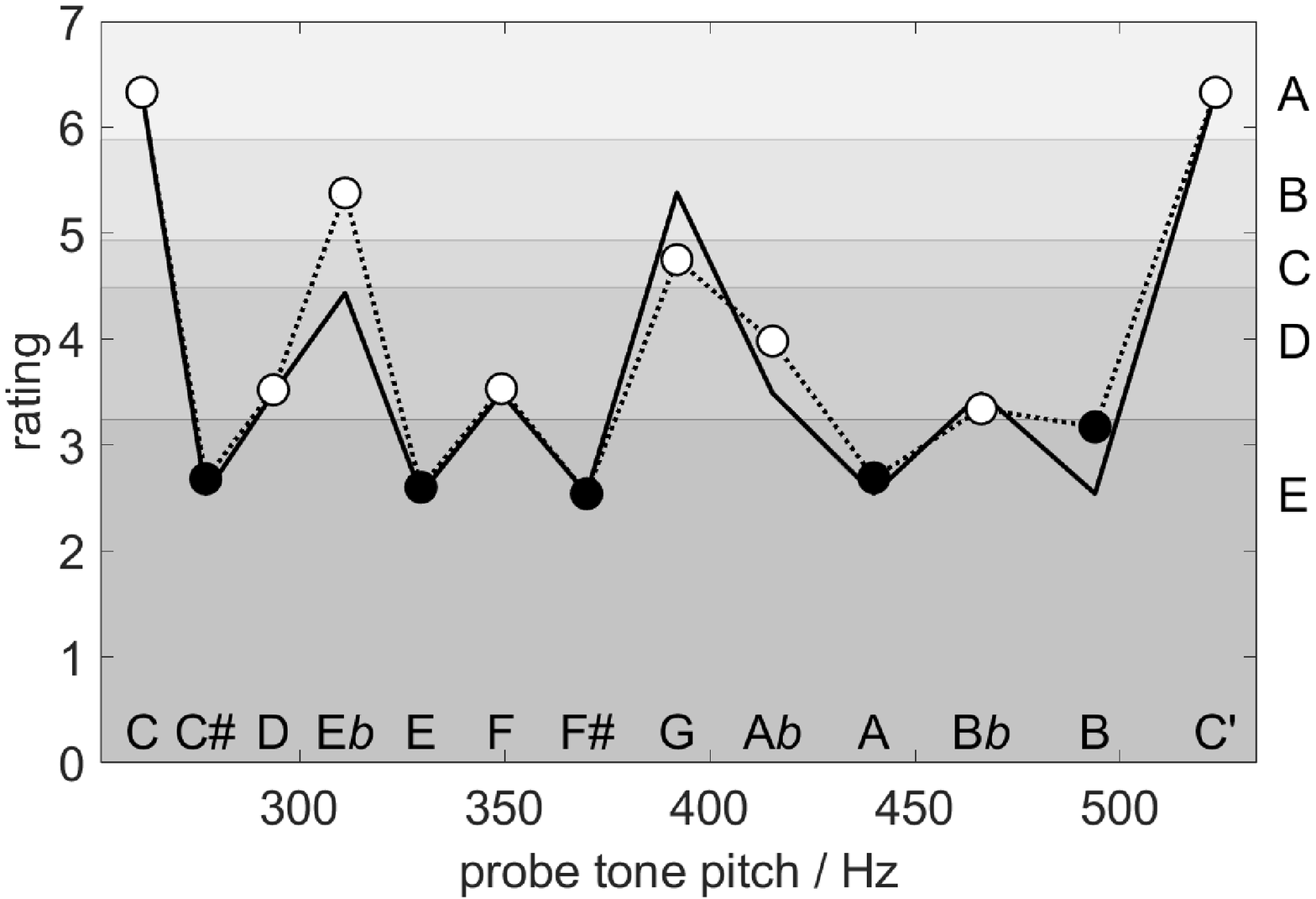}}
\caption{\label{fig:KK82frq} Static tonal attraction. Dotted with bullets: rating data $A_\mathrm{KK}(x)$ of \citet{KrumhanslKessler82} against physical pitch frequency $f$ in Hz,  solid: predictions from the hierarchical model [\Eq{eq:hierar}]. The labels A -- E at the right hand side indicate the labels of the hierarchical model [\Tab{tab:hiermod}]. (a) For C major context. (b) For C minor context. Open bullets indicate scale (diatonic) tones; closed bullets denote non-scale (chromatic) tones, for the respective scale.}
\end{figure}

The empirical results of \citet{KrumhanslKessler82} are replicated in \Fig{fig:KK82frq} as dotted curves connecting open and closed bullets. The probe tones are given as physical pitch frequencies (in Hz) at the $x$-axis. The subjective attraction ratings $A_\mathrm{KK}(x)$ (referring to \citeauthor{KrumhanslKessler82}) are plotted at the $y$-axis. Figure \ref{fig:KK82frq}(a) shows the results for the C major context, and \Fig{fig:KK82frq}(b) for the C minor context. The results of this experiment clearly show a kind of hierarchy: all diatonic scale tones (open bullets) received higher ratings than the chromatic nonscale tones (closed bullets). Moreover, in both modes, C major and C minor, the tonic C is mostly attractive, followed by the fifth, G and the third E for C major [\ref{fig:KK82frq}(a)] and by the third E$\flat$ and then the fifth G for C minor [\ref{fig:KK82frq}(b)].

Carrying out a multiscale analysis of their behavioral data, \citet{KrumhanslKessler82} recovered a geometric representation of keys akin to the third torus, thereby supporting the hierarchical models of tonal space as discussed in \Sec{sec:hier} \citep{Lerdahl88, Lerdahl96, KrumhanslKessler82, KrumhanslCuddy10}. We additionally present the predictions of the hierarchical model as the solid curves together with the labels of its levels in \Fig{fig:KK82frq}.

\subsection{Dynamic tonal attraction}
\label{sec:mmdyn}

Using five different chords, also consisting of Shepard tones, the C major triad CEG, the C minor triad CE$\flat$G, the dominant seventh CEGB$\flat$, a French sixth CEG$\flat$B$\flat$, and a half-diminished seventh CE$\flat$G$\flat$B$\flat$, as contexts, \citet{Woolhouse09} conducted a dynamic attraction probe tone experiment, asking listeners, how well a chromatic probe tone was melodically completing or harmonically resolving the priming context.

The results of this study are replicated in \Sec{sec:redyn}, \Fig{fig:W09mod} for the five contexts. The dotted curves connecting open and closed bullets present the original results of the \citet{Woolhouse09} experiment, where probe tones are arranged along the circle of fifths  [\Fig{fig:quinten}(b)] as real numbers $x = j \pi / 6$ ($j \in \mathbb{Z}_{12}$, with C(0) $\cong$ C'(12) one octave higher) at the $x$-axis.

Figure \ref{fig:W09mod}(a) displays the results for the C major priming context. The highest rated tone is F which is consistent with the interpretation of the priming C major triad as the dominant of F major. Another plausible interpretation of the prime would be the subdominant of G major, thus predicting higher likelihood of probe tone G. Similar results could be expected for the C minor context shown in \Fig{fig:W09mod}(b). In fact, F receives a relatively large rating, however, outperformed by A$\flat$. Figure \ref{fig:W09mod}(c) displays the dominant seventh chord which is similar to the C major triad. The data for the French sixth presented in \Fig{fig:W09mod}(d) give highest ratings for F and B and lowest ratings for D and A$\flat$. Finally, the half-diminished seventh [\Fig{fig:W09mod}(e)] is most likely resolved by F, C$\sharp$ and B.

\section{Theory}
\label{sec:theo}

In this section we develop our quantum models of static and dynamic tonal attraction, based on the fundamental principles of octave equivalence, fifth similarity and the universal musical transposition symmetry.

\subsection{Static tonal attraction}
\label{sec:thstatic}

Before we enter into the derivation of two possible quantum models for static tonal attraction, we briefly review the influential symbolic model based on the generative theory of tonal music \citep{LerdahlJackendoff83}.

\subsubsection{Hierarchical model}
\label{sec:hier}

The hierarchical model of \citet{Lerdahl88, Lerdahl96, KrumhanslKessler82}, and \citet{KrumhanslCuddy10}, depicted in \Tab{tab:hiermod}, comprises five levels of symbolic significance.

\begin{table}[H]
  \centering
  \begin{tabular}{l*{13}{c}}
  \hline
  A: octave & C &  &  & &  &  & &  & &  &  &  & C' \\
  B: fifths & C &  &  &  &  &  &  & G & &  &  &  & C' \\
  C: triadic & C & &  & & E &  & & G &  &  & &  & C' \\
  D: diatonic & C & & D & & E & F & & G & & A & & B & C' \\
  E: chromatic & C & C$\sharp$ & D & E$\flat$ & E & F & F$\sharp$ & G & A$\flat$ & A & B$\flat$ & B & C' \\
  \hline
  $s(j)$ & 4 & 0 & 1 & 0 & 2 & 1 & 0 & 3 & 0 & 1 & 0 & 1 & 4 \\
  \hline
\end{tabular}
  \caption{\label{tab:hiermod} Hierarchical model of static tonal attraction for C major context after \citet{Lerdahl96, Lerdahl15}.}
\end{table}

At the lowest, chromatic level E, all 12 tones of the chromatic octave are included. At the next, the diatonic level D, only the diatonic scale degrees are represented. One level higher, at the triadic level C, the three tones composing the triad according to major and minor third steps along the third torus are present. Again, one level higher, only tonic and fifths comprise the fifths level B. At the highest octave level A, eventually only the tonic prevails. The levels A -- E are indicated in \Fig{fig:KK82frq} as four shades of grey.

For each chromatic scale degree that serves as probe tone in the \citet{KrumhanslKessler82} experiment, one simply counts the number of degrees that are commonly shared across levels A to D (omitting level E that is common for all tones). The resulting number, $s(j)$, for probe tone $j \in \mathbb{Z}_{12}$ can be related to an attraction probability \citep{Temperley08}
\begin{equation}\label{eq:hierar}
    p(j) = \frac{s(j)}{\sum_j s(j)}
\end{equation}
that is rescaled and plotted in \Fig{fig:KK82frq} above as solid lines and correlated with the \citet{KrumhanslKessler82} data in \Tab{tab:stacor} in \Sec{sec:restatic}.\footnote{
    For C minor we use the so-called \emph{natural} minor scale which is obtained from the transposition of C major along the circle of fifths. There are two other minor scales, called \emph{harmonic} and \emph{melodic} minor which differ from natural minor in additional one or two semitone steps in ascending or descending lines.
}

The predictions of the hierarchical model are in good agreement with the experimental data. However, \Fig{fig:KK82frq}(b) indicates a slight deviation for C minor: the predicted attraction rate for the fifths G is larger and that of the minor third E$\flat$ is lower than the respective measured rate. We address this issue in section \ref{sec:asymdef} below.

\subsubsection{Free quantum model}
\label{sec:free}

A more instructive representation of the \citet{KrumhanslKessler82} data can be obtained by plotting the ratings $A_\mathrm{KK}(x)$ along the circle of fifths \citep{GrabenBlutner17}. This is done in \Sec{sec:restatic}, \Fig{fig:KK82free}. Here, the probe tones are represented as real numbers $x = j \pi / 6$ ($j \in \mathbb{Z}_{12}$, with C(0) $\cong$ C'(12) one octave higher) at the $x$-axis corresponding to the radian angles at the unit circle $S^1 = \mathbb{R} / 2\pi \mathbb{Z}$ interpreted as the circle of fifths [\Fig{fig:quinten}(b)]. The subjective attraction ratings $A_\mathrm{KK}(x)$ are plotted at the $y$-axis. Figure \ref{fig:KK82free}(a) shows the results for the C major context, and \ref{fig:KK82free}(b) for the C minor context.

The reordered rating profiles in \Fig{fig:KK82free} now exhibit some kind of periodicity: probe tones that are close to the tonic C ($x=0$) at the circle of fifths receive higher attraction values than tones that are close to the tritone F$\sharp$ ($x=\pi$). Therefore, the tritone F$\sharp$ may be seen as ``orthogonal'' to the tonic C in an appropriately chosen metric (cf. \citet{MannoneCompagno13, MannoneMazzola15} for alternative similarity assessments in musicology).

As a psychologically plausible model for static tonal attraction we therefore suggest the cosine similarity between the tonic context $0$ and a probe tone $x$,
\begin{equation}\label{eq:cosim}
    p(x) = \cos^2 \frac{x}{2}
\end{equation}
which directly reflects the auditory fifth similarity represented by the circle of fifths.

Cosine similarity is closely related to quantum similarity \citep{PothosBusemayerTrueblood13, PothosTrueblood15} in the framework of quantum cognition models \citep{BusemeyerBruza12, PothosBusemeyer13, BlutnerGraben16}. Therefore, we express the attraction value $p(x)$ through a continuous ``wave function'' $\psi(x)$ upon the circle of fifths which constitutes the one-dimensional ``tonal configuration space'' of our quantum model.\footnote{
    Note that the unit circle $S^1$ is a one-dimensional manifold, parameterized by a single continuous variable $x \in [0, 2\pi[$, that is  embedded into two-dimensional Euclidian space.
}

Quantum mechanical wave functions are particular instances of \emph{state vectors} in quantum theory. A state vector comprises a complete description of the state of a quantum system. It is contained in a vector space, called Hilbert space, that is equipped with a scalar product, thus allowing the calculation of projections. A quantum mechanical measurement device defines an orthogonal basis, such that the projections of a given state vector onto the individual basis vectors are interpreted as probabilities of the respective measurement results. Accordingly is the similarity of two quantum states given through the projection of one vector onto the other one, i.e. their common scalar product. Moreover, rotating a state vector in abstract Hilbert space must not change its physically relevant properties, i.e. projection probabilities. Therefore such rotations appear as \emph{symmetries} in quantum theory.

Our \emph{free quantum model} \citep{GrabenBlutner17} for the tonic wave function is hence
\begin{equation}\label{eq:statwav}
     \psi(x) = \frac{1}{\sqrt{\pi}} \cos \frac{x}{2} \:.
\end{equation}

The quantum wave function for an arbitrary context tone $a$ is obtained by applying a transposition operator $T_a$ to the tonic context $0$ through
\begin{equation}\label{eq:transpo1}
    \psi_a(x) = T_a \psi(x) = \psi(x - a)
\end{equation}
for the amount $a \in \mathbb{Z}_{12}$ rotating along the circle of fifths in clockwise direction. Applying the transposition operator to the tonic wave function \pref{eq:statwav}, we obtain
\begin{equation}\label{eq:deftra}
    \psi_a(x) = \frac{1}{\sqrt{\pi}} \cos \frac{x - a}{2}
\end{equation}
which can be expanded by virtue of the trigonometric addition theorems:
\begin{eqnarray*}
    \psi_a(x) &=& \frac{1}{\sqrt{\pi}} \cos \frac{x - a}{2} \\
    &=& \frac{1}{\sqrt{\pi}} \cos \frac{a}{2} \cos \frac{x}{2}  + \frac{1}{\sqrt{\pi}} \sin \frac{a}{2} \sin \frac{x}{2} \\
    &=& \frac{1}{\sqrt{\pi}} \cos \frac{a}{2} \cos \frac{x}{2}  + \frac{1}{\sqrt{\pi}} \sin \frac{a}{2} \cos \frac{x - \pi}{2} \\
    &=& \cos \frac{a}{2} \psi_0(x)  + \sin \frac{a}{2} \psi_\pi(x) \:.
\end{eqnarray*}
Hence any transposed state becomes a linear combination of two basic context states, the tonic (C) $\psi_0$, and the orthogonal tritone (F$\sharp$) $\psi_\pi$. The underlying Hilbert space of the free quantum model is therefore two-dimensional which proves the equivalence of this model with the earlier qubit model of \citet{Blutner15}.

The family $\{\psi_0,  \psi_\pi \}$ constitutes an orthonormal basis with respect to the wave function scalar product
\begin{equation}\label{eq:scpr}
    \langle \psi_y | \psi_a \rangle = \int_0^{2\pi} \psi_y(x)^* \psi_a(x) \D x
\end{equation}
where complex conjugation can be essentially neglected in the present case. Inserting the respective transposed wave functions yields
\begin{eqnarray*}
    \langle \psi_y | \psi_a \rangle &=&
    \Big\langle \cos \frac{y}{2} \psi_0  + \sin \frac{y}{2} \psi_\pi \Big| \cos \frac{a}{2} \psi_0  + \sin \frac{a}{2} \psi_\pi \Big\rangle \\
    &=&
    \cos \frac{y}{2} \cos \frac{a}{2} \langle \psi_0 | \psi_0 \rangle +
    \cos \frac{y}{2} \sin \frac{a}{2} \langle \psi_0 | \psi_\pi \rangle +
    \sin \frac{y}{2} \cos \frac{a}{2} \langle \psi_\pi | \psi_0 \rangle +
    \sin \frac{y}{2} \sin \frac{a}{2} \langle \psi_\pi | \psi_\pi \rangle \\
    &=&
    \cos \frac{y}{2} \cos \frac{a}{2} + \sin \frac{y}{2} \sin \frac{a}{2} \\
   &=& \cos \frac{y - a}{2}
\end{eqnarray*}
which is, up to a scaling factor,
\begin{equation}\label{eq:scprdef}
    \langle \psi_y | \psi_a \rangle = \sqrt{\pi} \psi(y - a) \:,
\end{equation}
the original wave function for the interval between probe tone $y$ and context tone $a$. Therefore, we can use the (scaled) quantum probability density
\begin{equation}\label{eq:qprob}
    p(x) = |\psi(x)|^2 = \psi(x)^* \psi(x)
\end{equation}
for a given interval as a measure of tonal attraction $p(y - a)$.

Sofar, our wave functions yield static tonal attraction values through the projection \pref{eq:scprdef} of a probe tone state onto a context tone state. In order to describe chord contexts, \citet{Woolhouse09, WoolhouseCross10} and \citet{Blutner15} suggested either to sum or to average the attraction profiles of individual pairs of tones over all pairings. For a context of two tones $a, b$ this \emph{Woolhouse conjecture} yields
\begin{equation}\label{eq:wool}
    p_{ab}(x) = |\psi_{ab}(x)|^2 = \frac{1}{2} \left( |\psi_{a}(x)|^2 + |\psi_{b}(x)|^2 \right) \:,
\end{equation}
in analogy to the density matrix formalism of statistical quantum mechanics (cf. \citet{Mannone18a} for a related account in musicology). For an arbitrary number $N$ of equally weighted context tones in a chord, we get
\begin{equation}\label{eq:wool2}
    p_C(x) = \frac{1}{N} \sum_{i = 1}^N |\psi_{a_i}(x)|^2 = \frac{1}{N} \sum_{i = 1}^N |\psi(x - a_i)|^2
\end{equation}
with context $C = \{ a_i \in \mathbb{Z}_{12} | i = 1, \dots, N\}$. When context tones are to be considered as differently weighted, we introduce weight factors $\rho(a_i)$ \citep{Mazzola02} and obtain a discrete convolution
\begin{equation}\label{eq:wool3}
    p_C(x) = \sum_{i = 1}^N \rho(a_i) p(x - a_i)
\end{equation}
with kernel $p(x) = |\psi(x)|^2$.

Let us examine the behavior of the discrete convolution under transposition $a \in \mathbb{Z}_{12}$. Applying the linear transposition operator $T_a$ to \pref{eq:wool3} yields
\begin{multline*}
    T_a p_C(x) = \sum_{i = 1}^N \rho(a_i) T_a  p(x - a_i) = \sum_{i = 1}^N \rho(a_i) p(x - a_i - a) = \\
    = \sum_{i = 1}^N \rho(a_i) p(x - (a_i + a) ) = \sum_{i = 1}^N \rho(b_i - a) p(x - b_i)
    = \sum_{i = 1}^N \rho(b_i) p(x - b_i) = p_{T_a C}(x)    \:,
\end{multline*}
if the convolution weights are transposition invariant
\begin{equation}\label{eq:coninv}
    \rho(b_i - a) = \rho(b_i)
\end{equation}
under the substitution $b_i = a_i + a$. Therefore, discrete convolution \pref{eq:wool3} implies transposition invariance \pref{eq:coninv}, or, in other words: transposition invariance is a necessary condition for the mixture of quantum states over chord contexts. This means that all context tones in a chord $C$ become transposed by the same interval $a \in \mathbb{Z}_{12}$: $T_a C = \{ b_i  \in \mathbb{Z}_{12} | b_i = a_i + a, i = 1, \dots, N\}$. This is definitely the case for equally weighted contexts $\rho(a_i) = 1/N$ used in our present exposition.\footnote{
    A more rigorous proof by means of harmonic analysis, which also entails the quantization of the continuous pitch spectrum into the discrete circle of fifths, will be published elsewhere.
}
In our quantum model, tones appear hence as wave fields over the circle of fifths, or, likewise, as \emph{Gestalts} in the sense of \citet{Kohler69}.

\subsubsection{Quantum deformation model}
\label{sec:def}

Our free quantum model from \Sec{sec:free} demonstrates that the tonal attraction data of the \citet{KrumhanslKessler82} experiment as shown in \Sec{sec:restatic}, \Fig{fig:KK82free} can be described in first approximation by a psychologically motivated cosine similarity model upon the circle of fifths. The structure of this geometrically represented tonal space is universal for music cognition as resulting from the fundamental principles of octave equivalence and transposition symmetry \citep{Lerdahl88, KrumhanslKessler82, Janata07}. In this subsection, we show how a straightforward modification of the free model leads to a more parsimonious representation of tonal space than the hierarchical model discussed in \Sec{sec:hier}.

To this aim, we introduce a suitable deformation \citep{GrabenBlutner17} of the distances along the circle of fifths by making the ansatz
\begin{equation}\label{eq:defowave}
     \psi(x) = A \cos (\gamma(x))
\end{equation}
for the stationary wave function where $\gamma(x)$ is a spatial deformation function and $A$ is a normalization constant.

\subsubsection{Symmetric deformation}
\label{sec:symdef}

In a first attempt we develop a symmetric deformation model for the \citet{KrumhanslKessler82} data. As a starting point we linearly rescale the empirical data
\begin{equation}\label{eq:scal}
    A_\mathrm{KK}(x) =  a \, p(x) + b \:.
\end{equation}
Inserting scaling constants $a = A_{\max} - A_{\min}$, $b = A_{\min}$, where $A_{\min}$ and $A_{\max}$ are the smallest and largest data values, and \Eq{eq:defowave}, and subsequently solving for $\gamma$ yields
\begin{equation}\label{eq:kkfit}
    \gamma(x) = \arccos \sqrt{ \frac{A_\mathrm{KK}(x) - A_{\min} }{ A_{\max} - A_{\min} } } \:.
\end{equation}

\begin{figure}[H]
\centering
\subfigure[]{\includegraphics[scale=0.3]{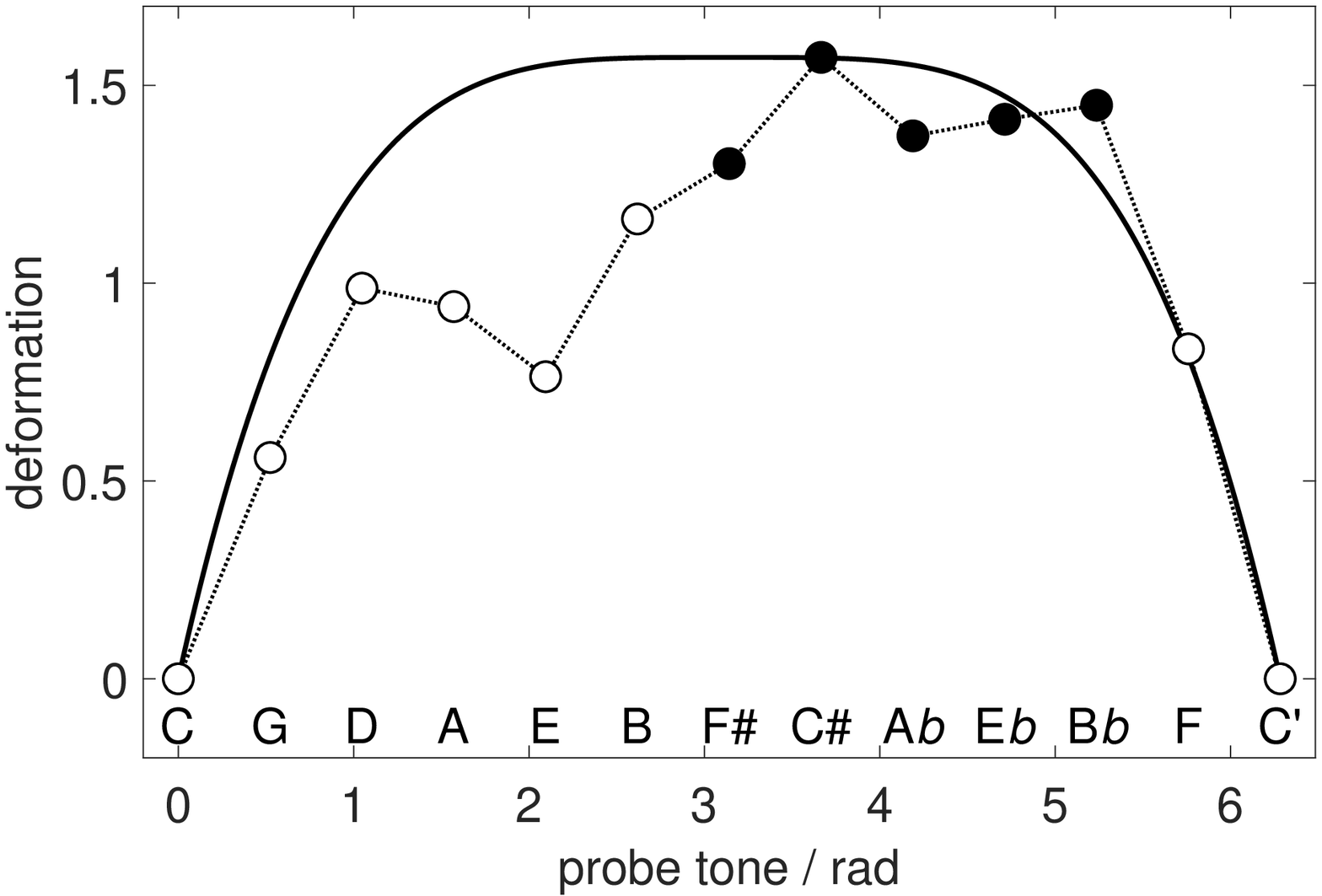}} \qquad
\subfigure[]{\includegraphics[scale=0.3]{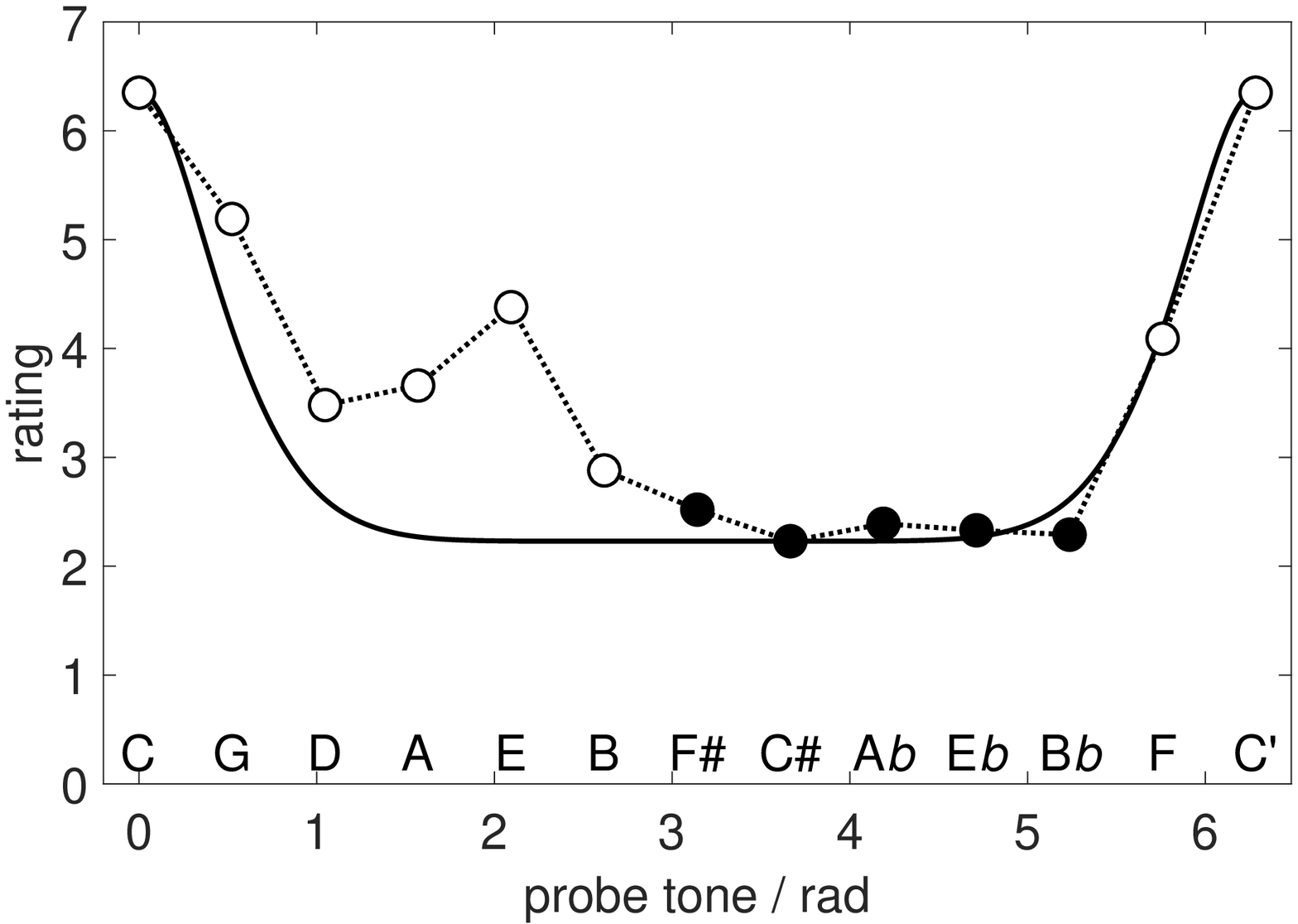}}
\caption{\label{fig:KK82fit} Optimal deformation of the cosine similarity profile for static tonal attraction. (a) Transformed \citet{KrumhanslKessler82} data [\Eq{eq:kkfit}] (dotted) against radian angles at the circle of fifths and deformation function \pref{eq:poly2} (solid). (b) Rating data $A_\mathrm{KK}(x)$ of \citet{KrumhanslKessler82} (dotted) and quantum probability kernel \Eq{eq:qprob} (solid). Both for C major context.}
\end{figure}

When we transform the C major data according to \pref{eq:kkfit} we obtain $\gamma(x)$ as the dotted curve in \Fig{fig:KK82fit}(a) resembling a buckled parabola (though neglecting the peak at E). Thus, we may fit the exponent $n$ of a symmetric $n$-th order polynomial
\begin{equation}\label{eq:poly}
    \gamma(x) = a_0 + a_n(x - \pi)^n \:,
\end{equation}
to the transformed data. For $n = 2$ the fit is rather poor ($r = 0.88$) and not really able to reproduce the buckle in the transformed data set. The next fit $n = 4$ works considerably well ($r = 0.96$), while exponents of higher than fourth order ($n = 6, 8$) do not substantially improve our fits. Thus, our choice $n = 4$ fits the data most parsimoniously.

The two parameters $a_0, a_4$ are necessarily obtained from two interpolation equations
\begin{eqnarray*}
    \gamma(0) &=& 0 \\
    \gamma(\pi) &=& \frac{\pi}{2} \:,
\end{eqnarray*}
i.e. the tonic should not be deformed while the tritone receives maximal deformation, as reflected by \Fig{fig:KK82fit}(a). Note that we do not perform a least-square fit for the parameters by minimizing some error functional. The parameters are fixed by the constraint that the deformation function should vanish at C and assume a maximum at F$\sharp$. This leads to the parsimonious model
\begin{equation}\label{eq:poly2}
    \gamma(x) =  \frac{\pi}{2} - \frac{(x - \pi)^4}{2 \pi^3} \:.
\end{equation}

We plot the original \citet{KrumhanslKessler82} data and the quantum probability density kernel \pref{eq:qprob} of the deformation model \pref{eq:defowave} in \Fig{fig:KK82fit}(b) \citep{GrabenBlutner17}.

At this point, it is useful to ask for the connection between the deformation model and the hierarchical model.  As one sees from \Fig{fig:KK82fit}(b), the kernel function of static tonal attraction assigns the maximum value to the target tone C. The two neighbors on the circle of fifth, i.e., G and F, get an attraction value that is about half of it. The attraction values of all other tones is very low such that we can neglect them. Hence, when we construct the attraction profiles for a certain context given by a triad (say CEG), we get an approximate reconstruction of the hierarchic model. The three tones of the triad (CEG) get a very high value; C and G a bit higher than E because of the discrete convolution in \Eq{eq:wool3}. Next, the neighbors of the triadic tones (C'G,F vs. G'D,C vs. E'B,A) are all diatonic tones and get an attraction of about 50\%. Hence, we can account for all levels of the hierarchic model shown in \Tab{tab:hiermod} besides the octave level (resulting in 4 different degrees of attraction).\footnote{
    A more sophisticated analysis by means of harmonic analysis, that will be published elsewhere, reveals that the Hilbert space of our deformed quantum model is infinite-dimensional in contrast to the two-dimensional case of the free quantum model discussed in \Sec{sec:free}.
}


\subsubsection{Asymmetric deformation}
\label{sec:asymdef}

The mirror symmetry of the spatial deformation model against the tritone F$\sharp$ leads to crucial problems when accounting for the differences between major and minor modes. So far, we considered the conception of static and dynamic attraction only. However, in cognitive music theory some other basic conceptions have been discussed including the idea of graded consonance/dissonance. According to \citet{Parncutt89}, the degree of (musical) consonance of a chord is related to the distribution of potential root tones of a chord. Hereby, the root tone of a chord can be seen as the tone with maximal static attraction given the chord as musical context. In cases with a single, incisive root tone, the chord sounds more consonant than in cases where several root tones compete against each other. Formally, we express the degree of chordal consonance as the static attraction value $p_C(x)$ [\Eq{eq:wool3}] of the (root) tone with maximum attraction after normalizing the attraction profile (i.e., the attraction values of the 12 tones sum up to 1).

Recently, \citet{JohnsonKangLeong12} investigated chords including major triads (CEG), minor triads (CE$\flat$G), diminished triads (CE$\flat$G$\flat$), and augmented triads (CEG$\sharp$). Table \ref{tab:asym} in \Sec{sec:restatic} shows the empirical ratings of the chord's consonance. Clearly, the major chords exhibit the highest degree of consonance followed by the minor chords. Further, the diminished chords are ranked lower and, at the bottom, we (surprisingly) find the augmented chords.  It is not difficult to see that the hierarchic model and the symmetric deformation model predict the same degrees of consonance for major and minor chords.

In order to further improve the predictive power of our quantum deformation model, we consider an asymmetric deformation polynomial of fourth order
\begin{equation}\label{eq:aspoly}
    \gamma(x) = a_0 + a_1 (x - \pi) + a_2 (x - \pi)^2 + a_3 (x - \pi)^3 + a_4 (x - \pi)^4  \:,
\end{equation}
where both, the linear term with coefficient $a_1$ and the cubic term with coefficient $a_3$ break the mirror-symmetry against the tritone F$\sharp$. As above, we demand that the tonic $x = 0 \cong 2 \pi$ is not deformed. This leads to two interpolation equations
\begin{eqnarray*}
    \gamma(0) &=& 0 \\
    \gamma(2 \pi) &=& 0 \:,
\end{eqnarray*}
allowing the elimination of two coefficients, e.g. $a_3, a_4$:
\begin{equation}\label{eq:aspoly2}
    \gamma(x) = a_0 + a_1 (x - \pi) + a_2 (x - \pi)^2 - \frac{a_1}{\pi^2} (x - \pi)^3
    - \frac{a_0 + a_2 \pi^2}{\pi^4}  (x - \pi)^4  \:.
\end{equation}
The remaining coefficients receive a straightforward interpretation as intercept: $a_0$, skewness: $a_1$, and steepness: $a_2$ (for moderate values). Equation \pref{eq:aspoly2} describes the symmetric deformation \pref{eq:poly2} under the choice $a_0 = \pi / 2, a_1 = a_2 = 0$.

We fit both parameters $a_1$ and $a_2$, after inserting \pref{eq:aspoly2} with the default values above into the convolution \pref{eq:wool3}, on the C major data set of \citet{KrumhanslKessler82} and obtain $a_1 = -0.14, a_2 = 0.01 \quad (r = 0.982)$. Finally, we compare our results with those of \citet{JohnsonKangLeong12} for C major and C minor triads, and for diminished triads and augmented triads, respectively. The results are presented in section \ref{sec:rasymdef}, \Tab{tab:asym}.

\subsection{Dynamic tonal attraction}
\label{sec:thdyn}

In this subsection we develop a quantum model based on a deformation of cosine similarity along the circle of fifths for the dynamic tonal attraction experiment of \citet{Woolhouse09}. This model is motivated by Woolhouse' \citeyearpar{Woolhouse09, WoolhouseCross10} interval cycle proximity measure that is described in the next section \ref{sec:icp}. Note that dynamic attraction has also successfully been described by Bayesian statistical modeling and Gaussian Markov chains trained on large musical corpora \citep{Temperley07, Temperley08}.

\subsubsection{Interval cycle model}
\label{sec:icp}

In recent research, \citet{Woolhouse09, WoolhouseCross10}, and \citet{Woolhouse12} have proposed to explain dynamic tonal attraction in terms of interval cycles. The basic idea is that the dynamic attraction between two pitches is proportional to the number of times the interval spanned by the two pitches must be multiplied by itself to produce some whole number of octaves. Assuming twelve-tone equal temperament, the \emph{interval-cycle proximity (ICP)} of an interval $j \in \mathbb{Z}_{12}$ can be defined as the smallest positive number $n$ such that the product with the interval length (i.e. the number of semitone steps spanned by the interval) is a multiple of twelve (i.e. the maximal interval length). More formally, let $j \in \mathbb{Z}_{12}$ be an interval on the chroma circle. The ICP $n = \icp(j)$ is defined as the smallest integer $n \in \mathbb{N}$, such that
\begin{equation}\label{eq:icpdef}
    n j = 0 \mod 12
\end{equation}
i.e. for a given $j$ one looks for the actually smallest $m \in \mathbb{N}$ such that $n j = 12 m$, which which eventually gives
\begin{equation}\label{eq:icp}
    \icp(j) = \frac{12 m}{j} \:.
\end{equation}
The following \Tab{tab:icp} lists the interval-cycle proximities for all intervals spanned by a given length. For example, one can see that the ICP of the tritone ($j=6$) is 2 and the ICP of the fifth ($j=7$) is 12. This has the plausible consequence that, relative to a root tone, the fifth has higher tonal attraction than the tritone.

\begin{table}[H]
  \centering
  \begin{tabular}{l*{13}{c}}
  \hline
  interval $j$ & 0 & 1 & 2 & 3 & 4 & 5 & 6 & 7 & 8 & 9 & 10 & 11 & 12 \\
  $\icp(j)$ & 1 & 12 & 6 & 4 & 3 & 12 & 2 & 12 & 3 & 4 & 6 & 12 & 1 \\
  \hline
\end{tabular}
  \caption{\label{tab:icp} Interval cycle proximity (ICP) on chroma circle.}
\end{table}

Two particular values of ICP are worth to be mentioned. Relative to the root, the semitone step (the minor second) receives the highest possible $\icp(1) = 12$, while the full-tone step (the major second) assumes its half, $\icp(2) = 6$. This fundamental relation, preferring small intervals in melodic dynamics against larger transitions is substantial for Western music \citep{CurtisBharucha09, KrumhanslEA00, Narmour92, Temperley08} and should be maintained in any alternative proposal to the IC model, such as in our quantum model below.

ICP exhibits two interesting symmetries. First, it is obviously invariant under reflection against the tritone $j = 6$, therefore $\icp(j) = \icp(12 - j)$. Second, it is invariant under the transition from the chroma circle to the circle of fifths and vice versa \citep{WoolhouseCross10}: $\icp(j) = \icp(7 j \mod 12)$. As a consequence, ICP assigns the highest value 12 to the minor second ($j=1$) and to the forth ($j=5$). We may thus represent our wave functions alternatively either at the chroma circle, or at the circle of fifths. ICP are plotted along the chroma circle in \Fig{fig:W09fit}(b) as the dotted curve.

\subsubsection{Deformation model}
\label{sec:defm}

In order to construct a quantum deformation model of the dynamic tonal attraction data of \citet{Woolhouse09} we realize that a symmetric fourth-order polynomial is not sufficient to model ICP which we though consider a good phenomenal working model for the resolution of chords into single tones, despite ongoing discussions \citep{Quinn10}. Therefore, we chose the next possible model class of symmetric sixth-order polynomials. We try to interpolate
\begin{equation}\label{eq:polydy}
    \gamma(x) = a_0 + a_2 (x - \pi)^2 + a_4 (x - \pi)^4 + a_6 (x - \pi)^6
\end{equation}
to the transformed and rescaled ICP (cf. \Eq{eq:kkfit})
\begin{equation}\label{eq:icpfit}
    \gamma(x) = \arccos \sqrt{\frac{\icp(x) - 1}{11}} \:.
\end{equation}

Equation \pref{eq:icpfit} is ambiguous with respect to the sign of the square root. Taking only the positive square root of ICP into account, involves zeros of $\gamma(x)$ with multiplicities larger than one, thereby generating a highly fluctuating function which prevents interpolation by a sixth-order polynomial. Thus, we avoid this ambiguity by a smoothing function $\sigma(j)$ tabulated in \Tab{tab:mask}.

\begin{table}[H]
  \centering
  \begin{tabular}{l*{13}{c}}
  \hline
  interval $j$ & 0 & 1 & 2 & 3 & 4 & 5 & 6 & 7 & 8 & 9 & 10 & 11 & 12 \\
  $\sigma(j)$ & -1 & 1 & 1 & 1 & 1 & 1 & -1 & 1 & 1 & 1 & 1 & 1 & -1\\
  \hline
\end{tabular}
  \caption{\label{tab:mask} Smoothing function $\sigma$ forcing simple zeros of the ICP deformation.}
\end{table}

\begin{figure}[H]
\centering
\subfigure[]{\includegraphics[scale=0.3]{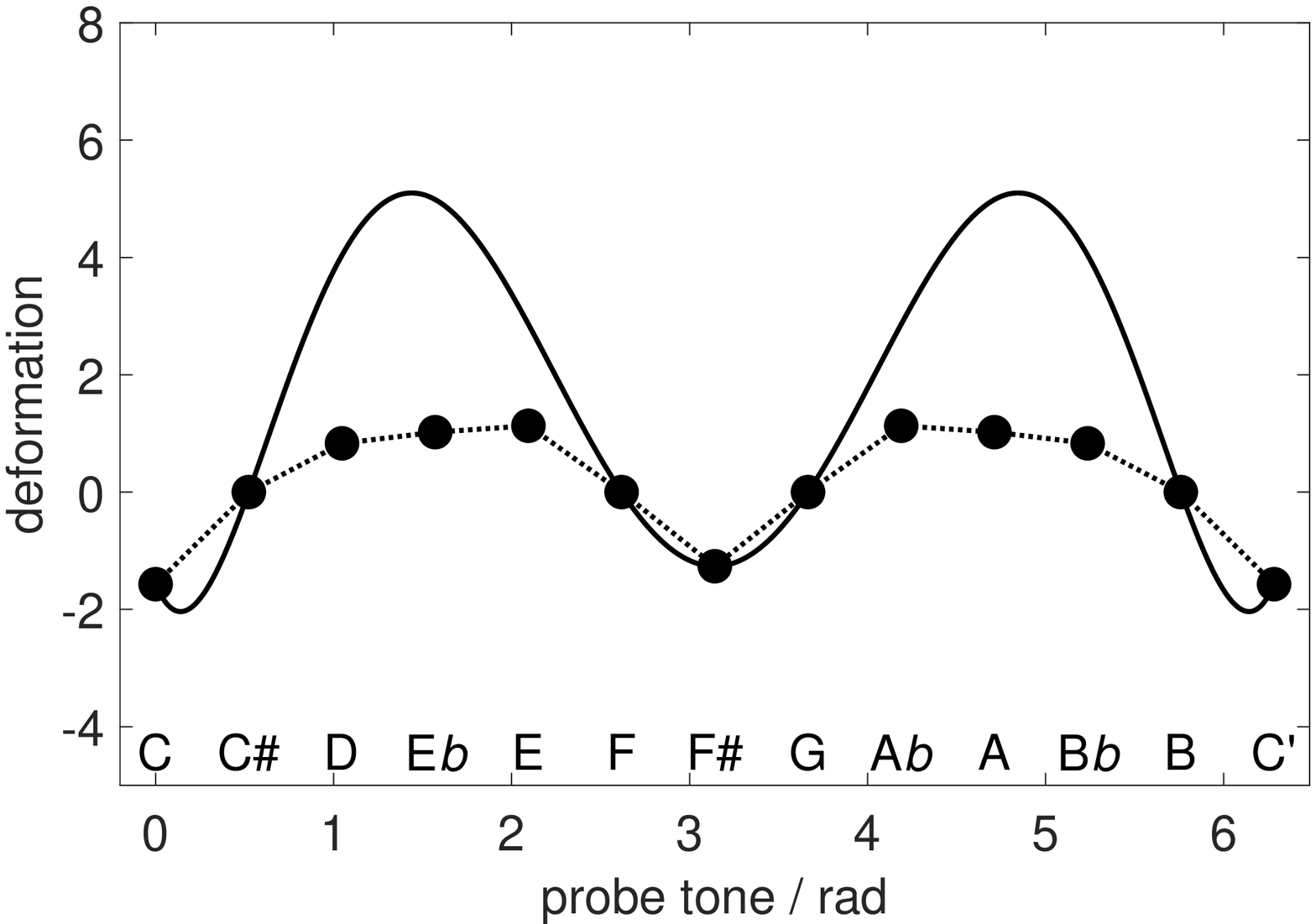}} \qquad
\subfigure[]{\includegraphics[scale=0.3]{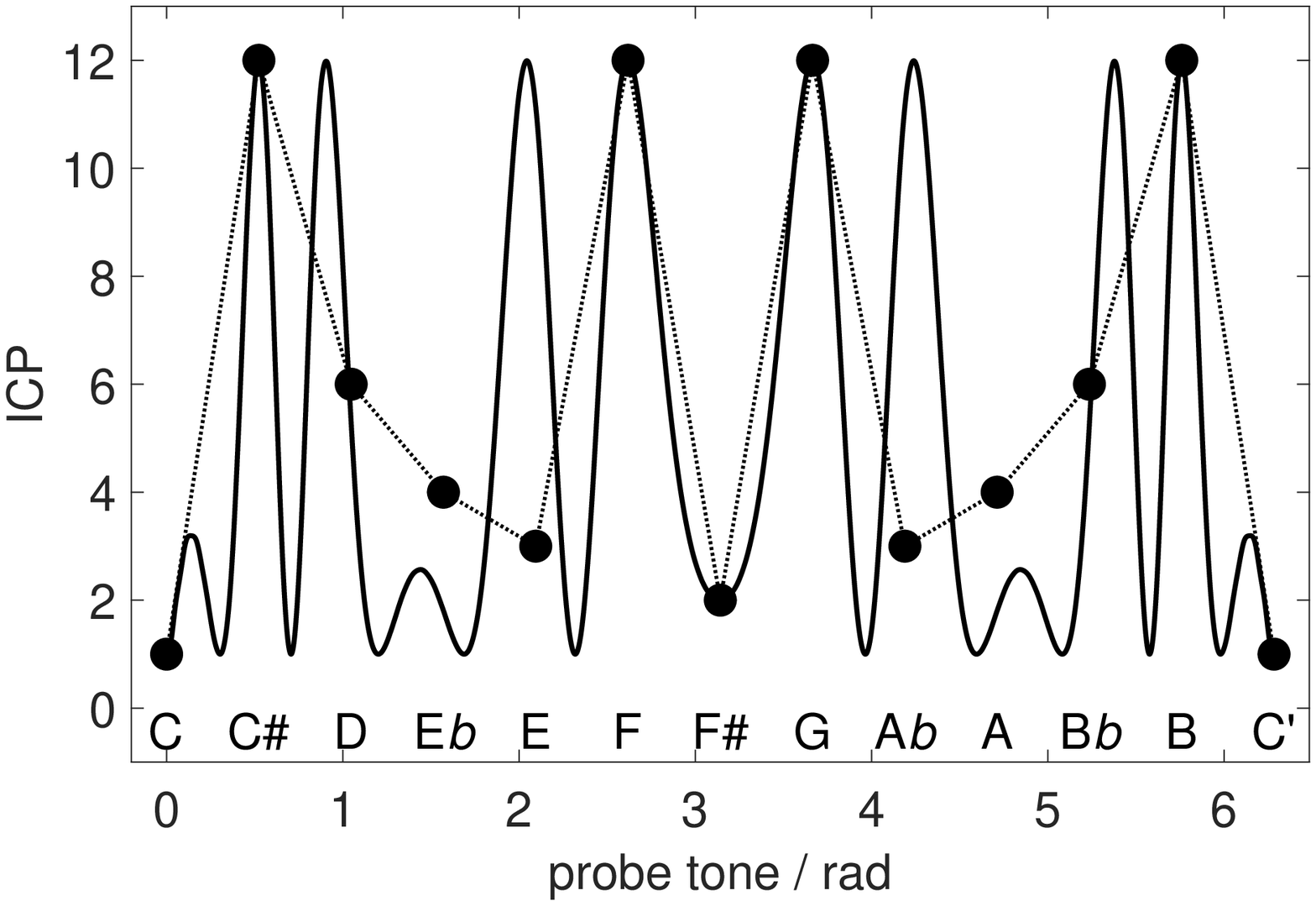}}
\caption{\label{fig:W09fit} Optimal deformation of the cosine similarity profile for dynamic tonal attraction. (a) Transformed ICP data \pref{eq:icpfit} along the chroma circle (dotted), deformation function \pref{eq:polydy2} (solid). (b) ICP (dotted) and quantum deformation model (solid).}
\end{figure}

The smoothing function $\sigma$ assigns negative values to the unison ($j=0$), to the tritone ($j=6$) and to the octave ($j=12$) and positive values to all other intervals. Consequently, the zeros of $\gamma(x)$ at the minor second ($j=1$) and the forth ($j=5$) become simple zeros, with positive slope for $j=1$ and negative slope for ($j=5$), such that the graph of $\gamma$ traverses the $x$-axis, thus reducing the complexity of the interpolation. For the values tabulated in \Tab{tab:mask}, $\gamma(x)$ can be interpolated with a symmetric fourth-order polynomial. Figure \ref{fig:W09fit}(a) shows the resulting function $\gamma(x) / \sigma(x)$ in black.

Instead of rigorously modeling the interval cycle model, we use it more as a guideline for our quantum model. Therefore, our model should at least share three important features with ICP: It should prefer small intervals $j = 1, 2$ and it should respect the symmetries of ICP. This choice already fixes two intervals for our interpolation, namely the zeros of the transformed ICP $\gamma$ at $j = 1, 5$, where the cosine similarity function \pref{eq:defowave} should not be deformed at all. For the third and fourth interpolation points, we select as in \Sec{sec:def} the tonic at $j=0$ and the tritone at $j=6$ as the symmetry center. Then, the four parameters $a_0, a_2, a_4, a_6$ are necessarily (without any fit) obtained from four interpolation equations
\begin{eqnarray*}
    \gamma(0) &=& -\frac{\pi}{2} \\
    \gamma\left(\frac{\pi}{6}\right) &=& 0 \\
    \gamma\left(\frac{5 \pi}{6}\right) &=& 0 \\
    \gamma(\pi) &=& -\arccos 11^{-\frac{1}{2}} \:.
\end{eqnarray*}
The result is the parsimonious model
\begin{multline}\label{eq:polydy2}
    \gamma(x) =  -\arccos 11^{-\frac{1}{2}}
    - \frac{125 \pi - 147994 \arccos 11^{-\frac{1}{2}}}{3850 \pi^2} (x - \pi)^2 + \\
    + \frac{2340 \pi - 171864 \arccos 11^{-\frac{1}{2}}}{1925 \pi^4} (x - \pi)^4
    - \frac{3240 \pi - 99792 \arccos 11^{-\frac{1}{2}}}{1925 \pi^6} (x - \pi)^6 \:.
\end{multline}

We plot the original ICP \pref{eq:icp} and the quantum probability density \pref{eq:qprob} of the deformation model \pref{eq:polydy2} \Fig{fig:W09fit}(b). Apparently, the kernel resulting from \pref{eq:polydy2} displays some fluctuations over the chroma circle. Interestingly, although the deformation was interpolated only at four intervals C, C$\sharp$, F and F$\sharp$ in \Fig{fig:W09fit}(a), the resulting attraction profile agrees with ICP also at the interval D which is crucial for predicting the ratings of the major second interval. However, or interpolation substantially deviates from ICP at minor and major thirds E$\flat$ and E.

\section{Results}
\label{sec:res}

This section summarizes the results of our quantum models for tonal attraction which will be compared with traditional models of \citet{Lerdahl88, Lerdahl96, KrumhanslKessler82} for the static and \citet{Woolhouse09, WoolhouseCross10} for the dynamic attraction data.

\subsection{Static tonal attraction}
\label{sec:restatic}

First, we consider a symmetric deformation function as derived in \Sec{sec:symdef}.


\subsubsection{Symmetric deformation}
\label{sec:resymdef}

\begin{figure}[H]
\centering
\subfigure[]{\includegraphics[scale=0.3]{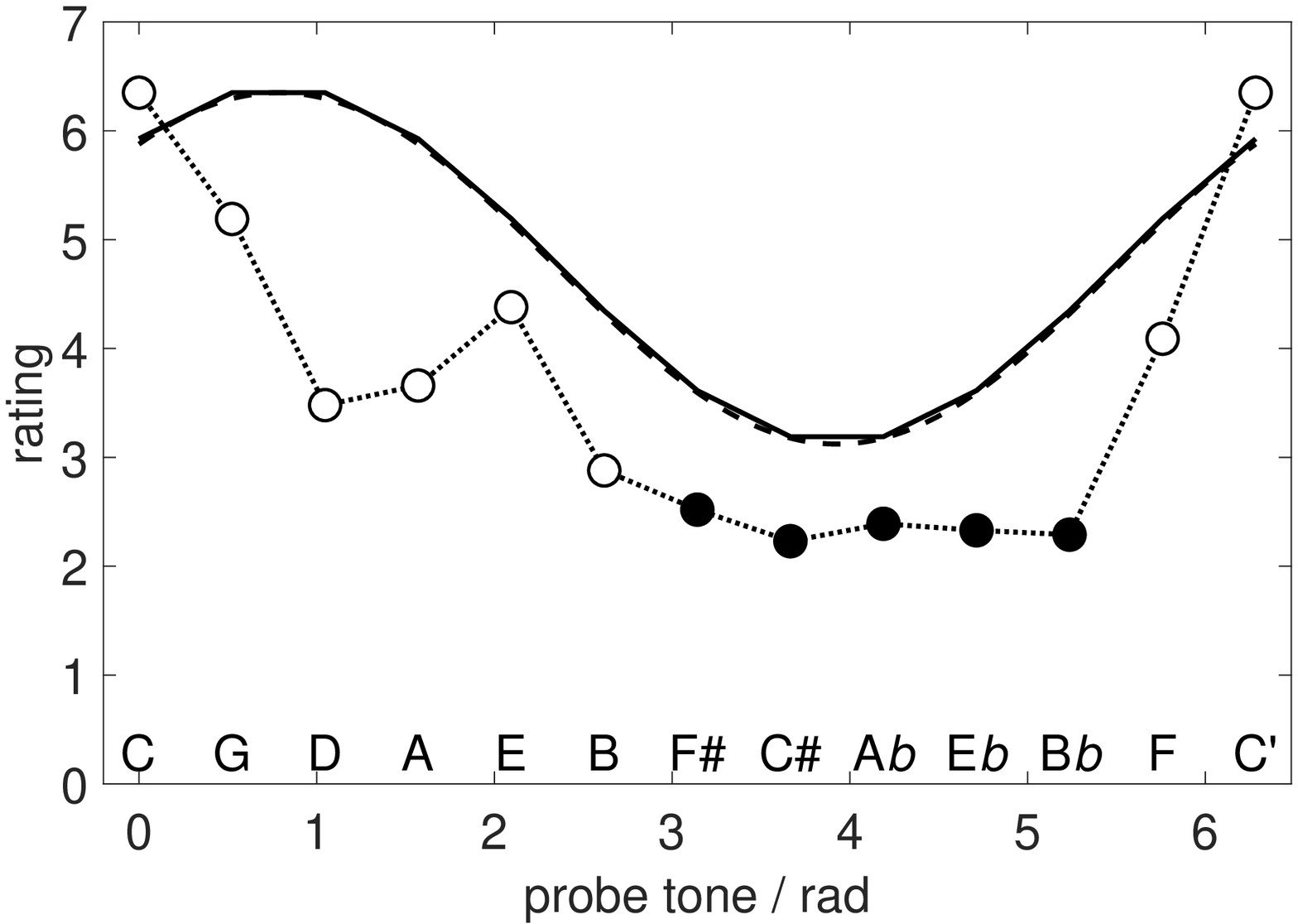}} \qquad
\subfigure[]{\includegraphics[scale=0.3]{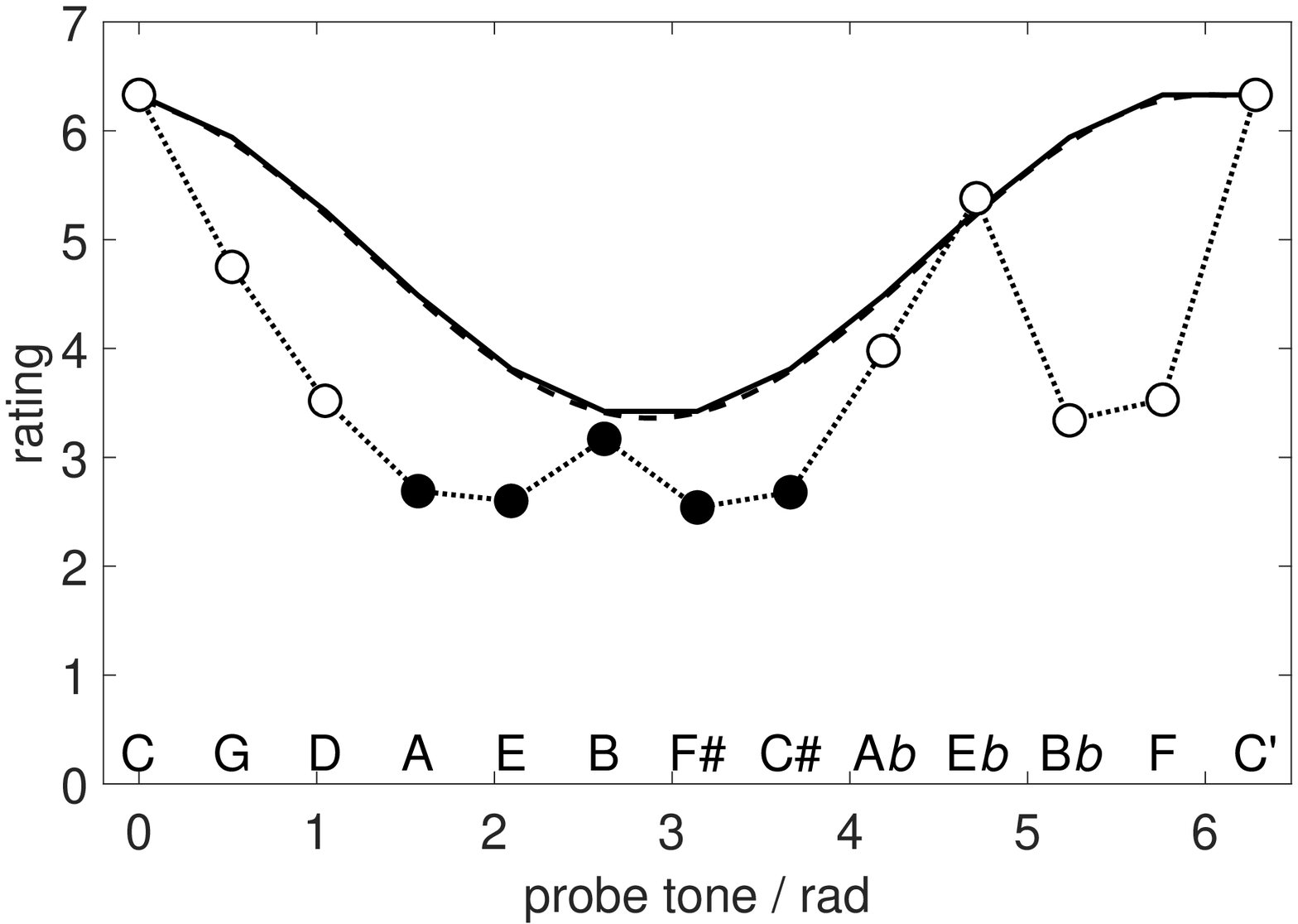}}
\caption{\label{fig:KK82free} Free quantum (cosine similarity) model for static tonal attraction. Dotted with bullets: rating data $A_\mathrm{KK}(x)$ of \citet{KrumhanslKessler82} against radian angles at the circle of fifths, dashed: continuous mixed state model $p_C(x)$ [\Eq{eq:wool3}]. (a) For C major context. (b) For C minor context. Solid: $p_C(x)$ after discrete sampling across the circle of fifths. Open bullets indicate scale (diatonic) tones; closed bullets denote non-scale (chromatic) tones.}
\end{figure}

Computing the quantum probabilities \pref{eq:qprob} for the free model gives the results in \Fig{fig:KK82free}. Figure \ref{fig:KK82free}(a) shows the agreement of the rescaled quantum probability densities with the \citet{KrumhanslKessler82} data for C major, and \Fig{fig:KK82free}(b) for C minor. The dotted curves display the results of the \citet{KrumhanslKessler82} experiment, arranged along the circle of fifths. The dashed curves show the continuous quantum probability density $p_C(x)$ [\Eq{eq:wool3}] obtained from the convolution of the kernel function $p(x)$ \pref{eq:qprob} of the free pure state wave function $\psi(x)$ [\Eq{eq:statwav}] over the major triad context CEG (a) and the minor triad context CE$\flat$G (b). The solid curves display $p_C(x)$ discretely sampled across the circle of fifths.

\begin{figure}[H]
\centering
\subfigure[]{\includegraphics[scale=0.3]{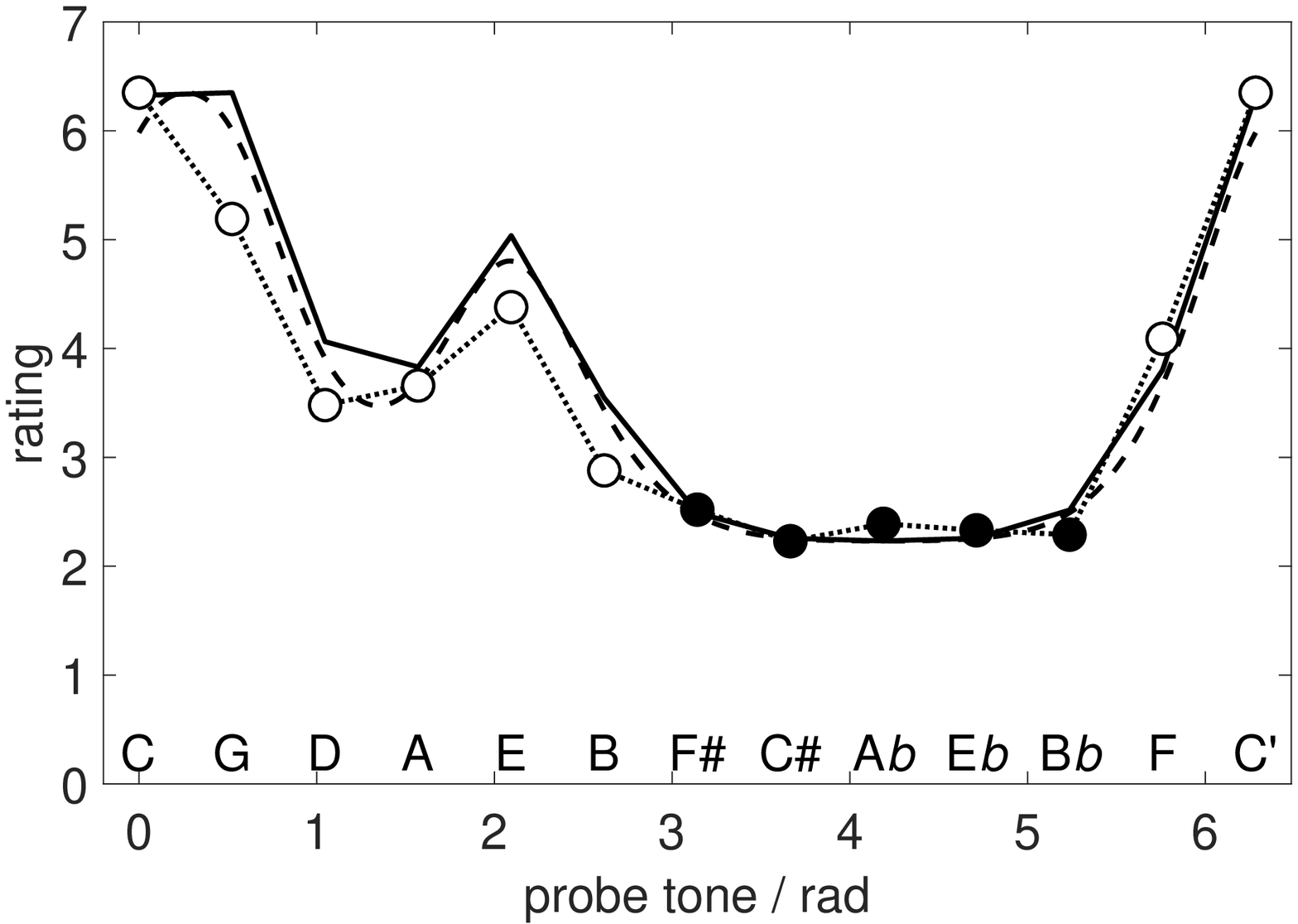}} \qquad
\subfigure[]{\includegraphics[scale=0.3]{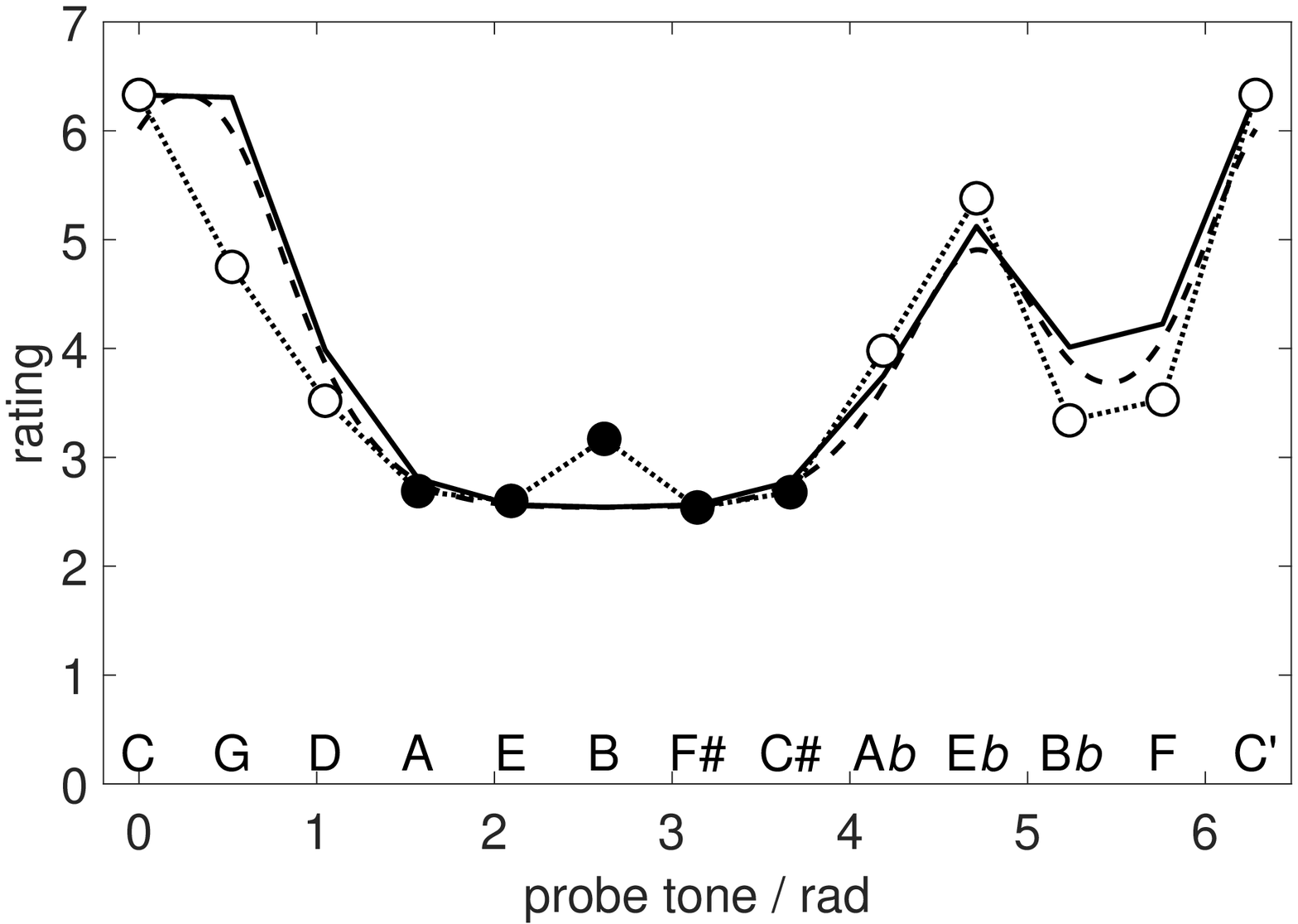}}
 \caption{\label{fig:KK82deform} Quantum deformation model for static tonal attraction. Dotted with bullets: rating data $A_\mathrm{KK}(x)$ of \citet{KrumhanslKessler82} against radian angles at the circle of fifths, dashed: mixed state model $p_C(x)$ [\Eq{eq:wool3}]. (a) For C major context. (b) For C minor context. Solid: $p_C(x)$ after discrete sampling across the circle of fifths. Open bullets indicate scale (diatonic) tones; closed bullets denote non-scale (chromatic) tones.}
\end{figure}

In \Fig{fig:KK82deform} we present the results of modeling the \citet{KrumhanslKessler82} data (dotted) with the quantum deformation approach [Eqs. \pref{eq:defowave}, \pref{eq:poly2}]. Computing the mixed state quantum probabilities $p_C(x)$ as convolutions of $p(x)$ [\pref{eq:qprob}] over the major triad context CEG (a) and the minor triad context CE$\flat$G (b) [\Eq{eq:wool3}] gives the dashed curves. The solid curves display $p_C(x)$ after discrete sampling.

Obviously, the mixed state quantum deformation model tracks the experimental data almost perfectly. In particular it is able to unveil the different levels of the hierarchical model: the tonic receives maximal attraction, followed by the fifth, by the third, and eventually by the remaining diatonic scale tones. However, the model does not capture the slight asymmetry between major and minor modes in the data.

Carrying out a correlation analysis between model and data yields the results in \Tab{tab:stacor}.

\begin{table}[H]
  \centering
  \begin{tabular}{l*{5}{c}} \hline
  {}& hierarchical model & \multicolumn{2}{c}{free model} & \multicolumn{2}{c}{sym. deformation model} \\
  {} {} & & pure  & mixed  & pure & mixed \\
  C major &  0.98 & 0.70 & 0.78 & 0.89 & 0.97 \\
  C minor &  0.95 & 0.68 & 0.72 & 0.80 & 0.93 \\
  \hline
  \end{tabular}
  \caption{\label{tab:stacor} Correlation coefficients $r$ of four discussed models for static tonal attraction with \citet{KrumhanslKessler82} data. All error probabilities are significantly below 0.05.}
\end{table}

The free cosine similarity model already accounts for about $r^2 = 50\%$  of the data's variance for pure quantum states and slightly improves for mixed quantum states convolved over contextual triads. The deformation model performs much better for pure quantum states and does as good as the hierarchical model based on the generative theory of tonal music. Hence, we conclude that our quantum model is able to explain the experimental results on static tonal attraction using a parsimonious approach.

This contrasts with the hierarchical model in methodological and empirical respects. As outlined in \Sec{sec:hier}, the hierarchical model stipulates all five levels of description that are essential for predicting tonal attraction ratings. In this sense, the hierarchical model lacks explanatory power. Our deformation model, however, only has to stipulate the triadic level of the hierarchical model. This level is used for calculating the discrete convolution \pref{eq:wool3} of the deformation kernel \pref{eq:qprob}. From this, the entire major and minor scales are rendered. As discussed in \Sec{sec:def}, the deformation ansatz derives all levels of the hierarchic model, besides the octave level, from the triadic context together with the mixing conjecture and does not require any further stipulation. Thus, our model parsimoniously outperforms the hierarchical model in terms of explanatory power.


\subsubsection{Asymmetric deformation}
\label{sec:rasymdef}

Finally, we present the results for the asymmetric deformation model from \Sec{sec:asymdef} in \Tab{tab:asym}. The asymmetric quantum model is able to explain the correct consonance ranking of major (CEG), minor (CE$\flat$G), diminished (CE$\flat$G$\flat$), and augmented triads (CEG$\sharp$), respectively \citep{JohnsonKangLeong12}. For the hierarchical model the correlation is $r = 0.93$ ($p = 0.07$), for the asymmetric deformation model we have $r = 0.95$ ($p = 0.05$).

\begin{table}[H]
  \centering
  \begin{tabular}{l*{3}{c}} \hline
  {} & emp. consonance  & hier. model & asym. deformation model \\
    major	& 5.33	& 0.49	& 0.54 \\
    minor	& 4.59	& 0.49	& 0.52 \\
    diminished	& 3.11	& 0.34	& 0.36 \\
    augmented	& 1.74	& 0.33	& 0.34 \\
  \hline
  \end{tabular}
  \caption{\label{tab:asym} Empirical consonance ratings and model predictions for common triads. The predictions of the models concern the strength of strongest static attraction using normalized attraction profiles.}
\end{table}

Summarizing, we consider a case of symmetry breaking in cognitive musicology, breaking the mirror symmetry of the tonal attraction kernel in analogy to \citet{Parncutt11}. In this way, we overcome some weaknesses of the classical attraction model based on tonal hierarchies, which cannot account for the differences between major and minor modes. Consequently, we get a suitable model not only for static attraction profiles but also for graded consonance/dissonance. The ability for unification --- grasping different phenomena in a systematic way --- is one of the trademarks of quantum theory, which is correspondingly apparent in the domain of quantum cognition as well.

\subsection{Dynamic tonal attraction}
\label{sec:redyn}

\begin{figure}[H]
\centering
\subfigure[]{\includegraphics[scale=0.3]{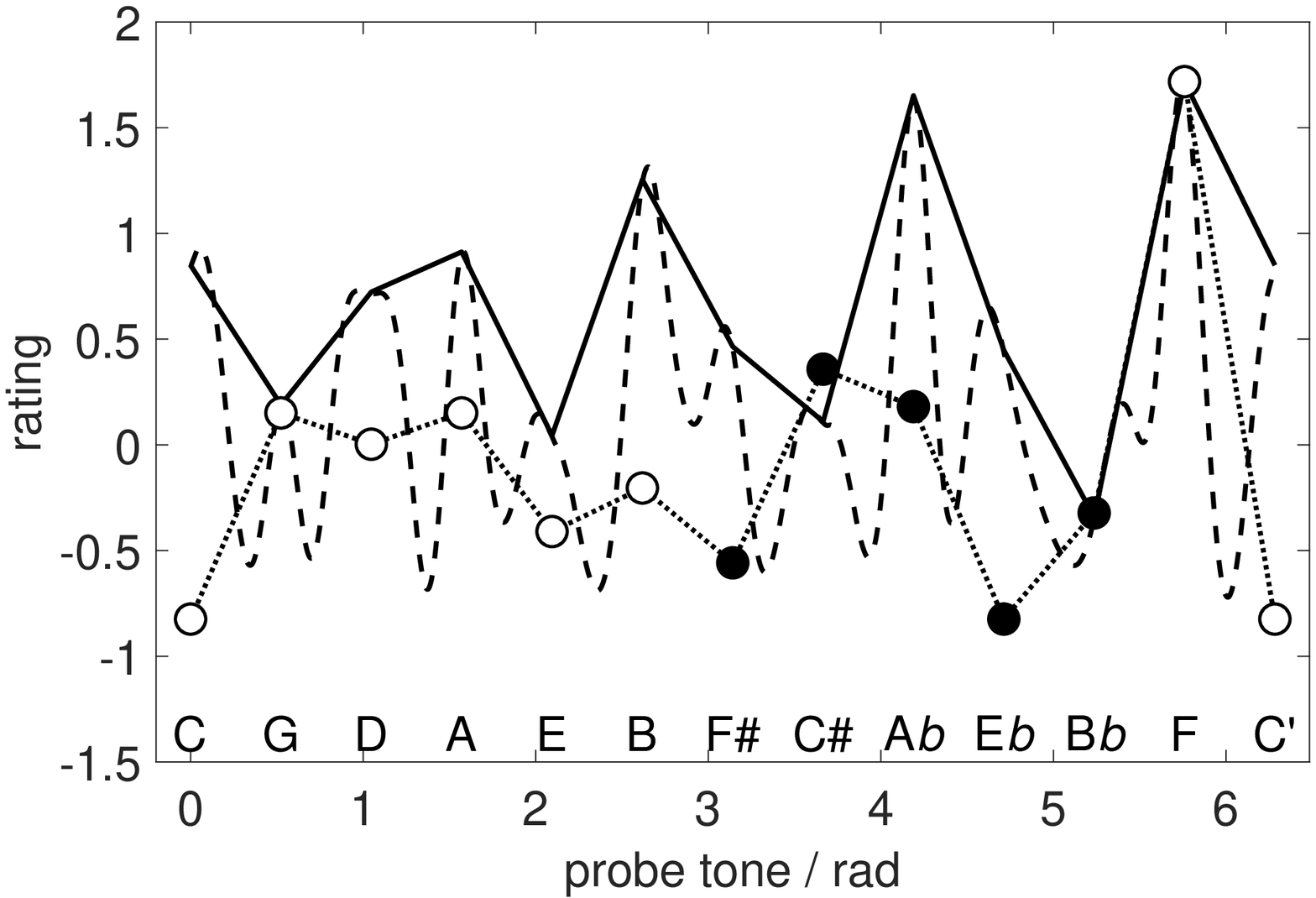}} \qquad
\subfigure[]{\includegraphics[scale=0.3]{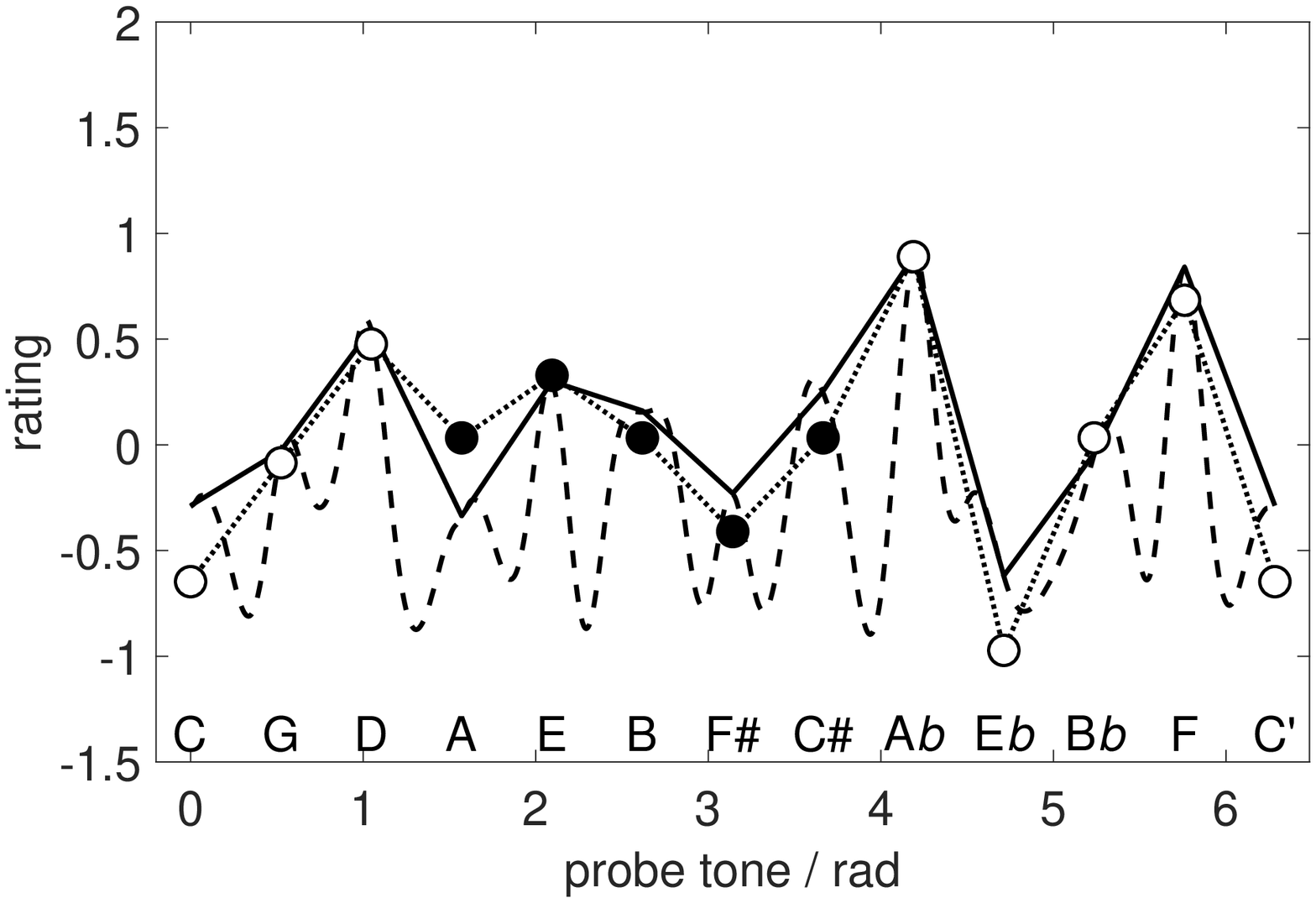}}
\subfigure[]{\includegraphics[scale=0.3]{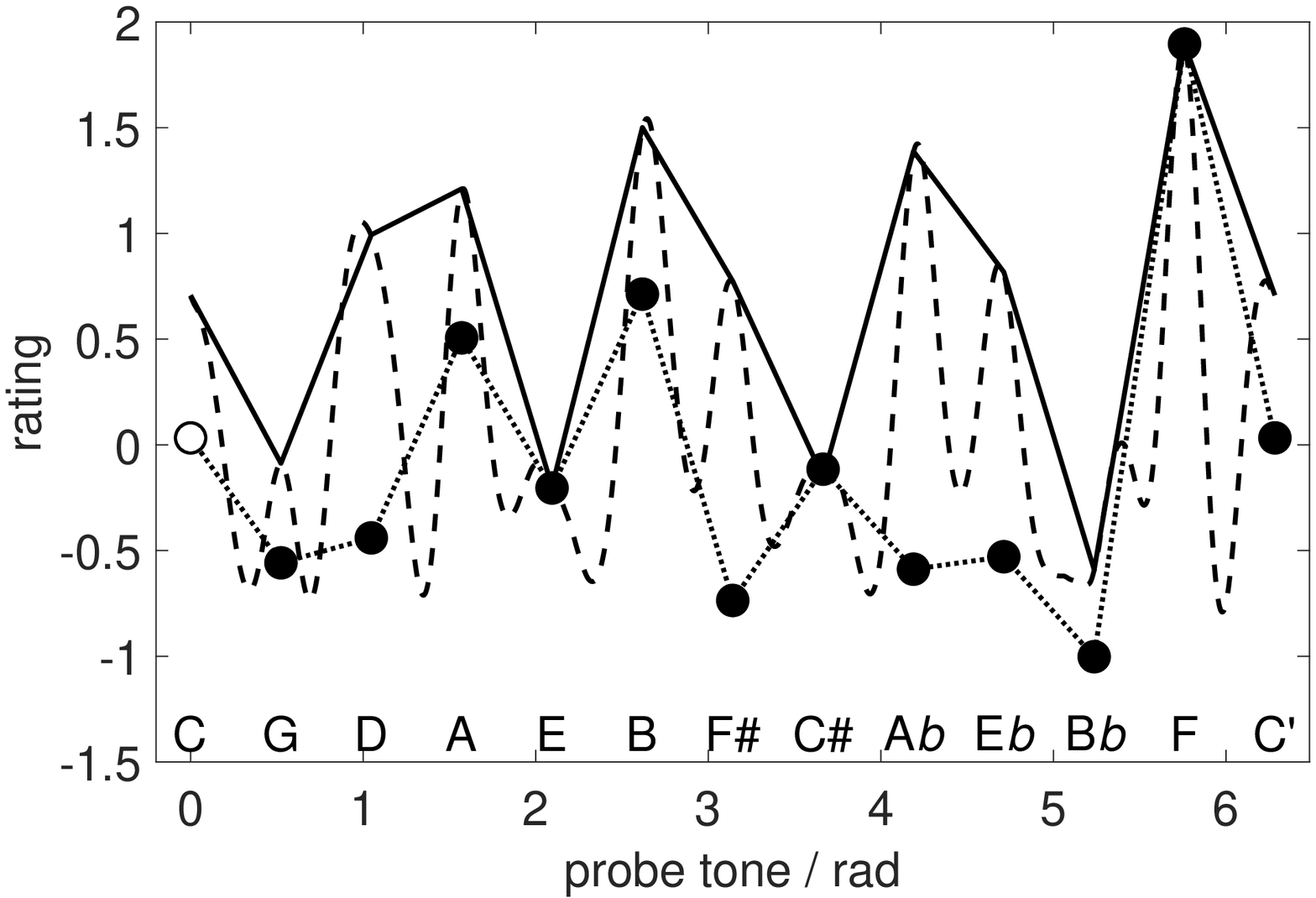}} \qquad
\subfigure[]{\includegraphics[scale=0.3]{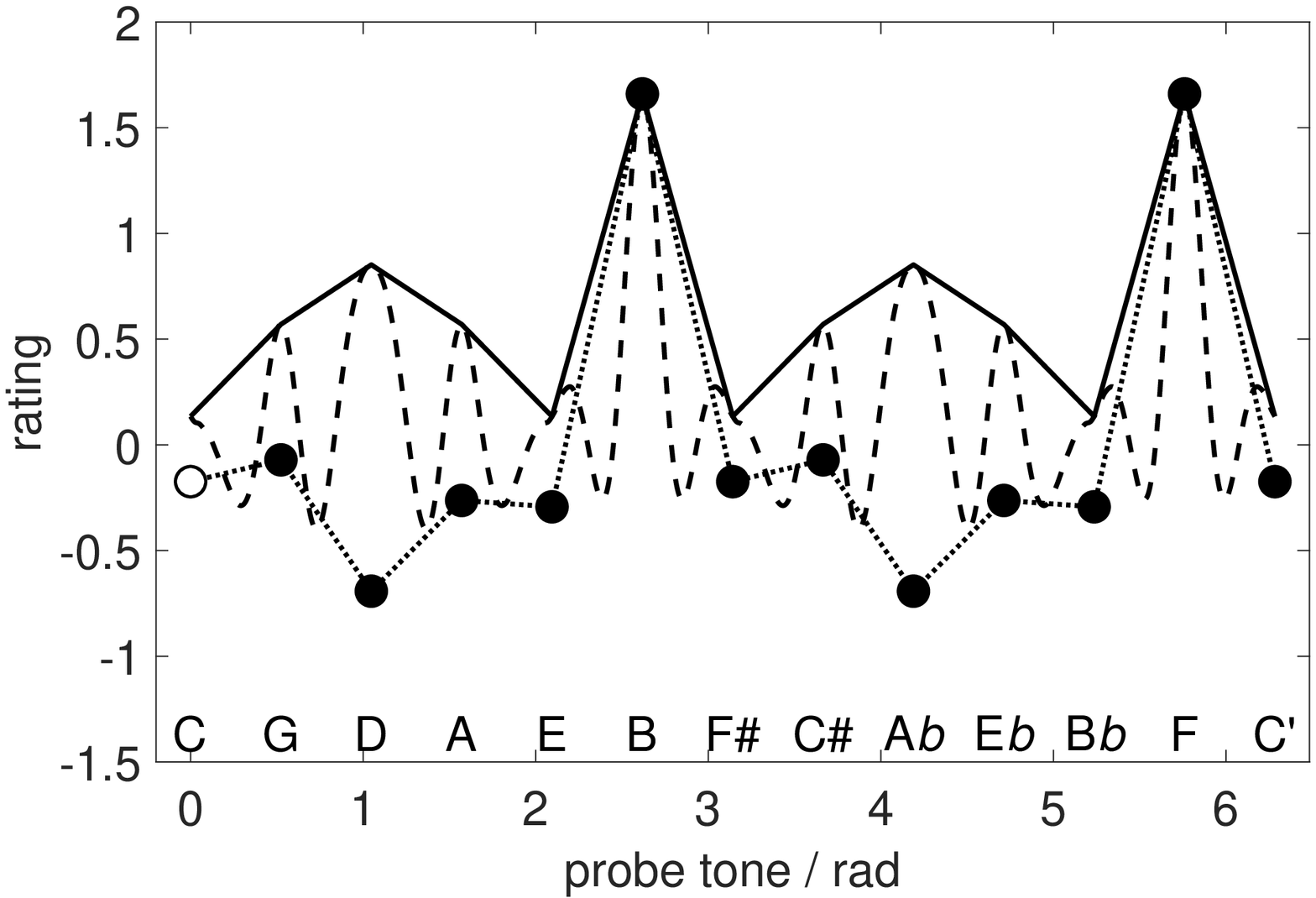}}
\subfigure[]{\includegraphics[scale=0.3]{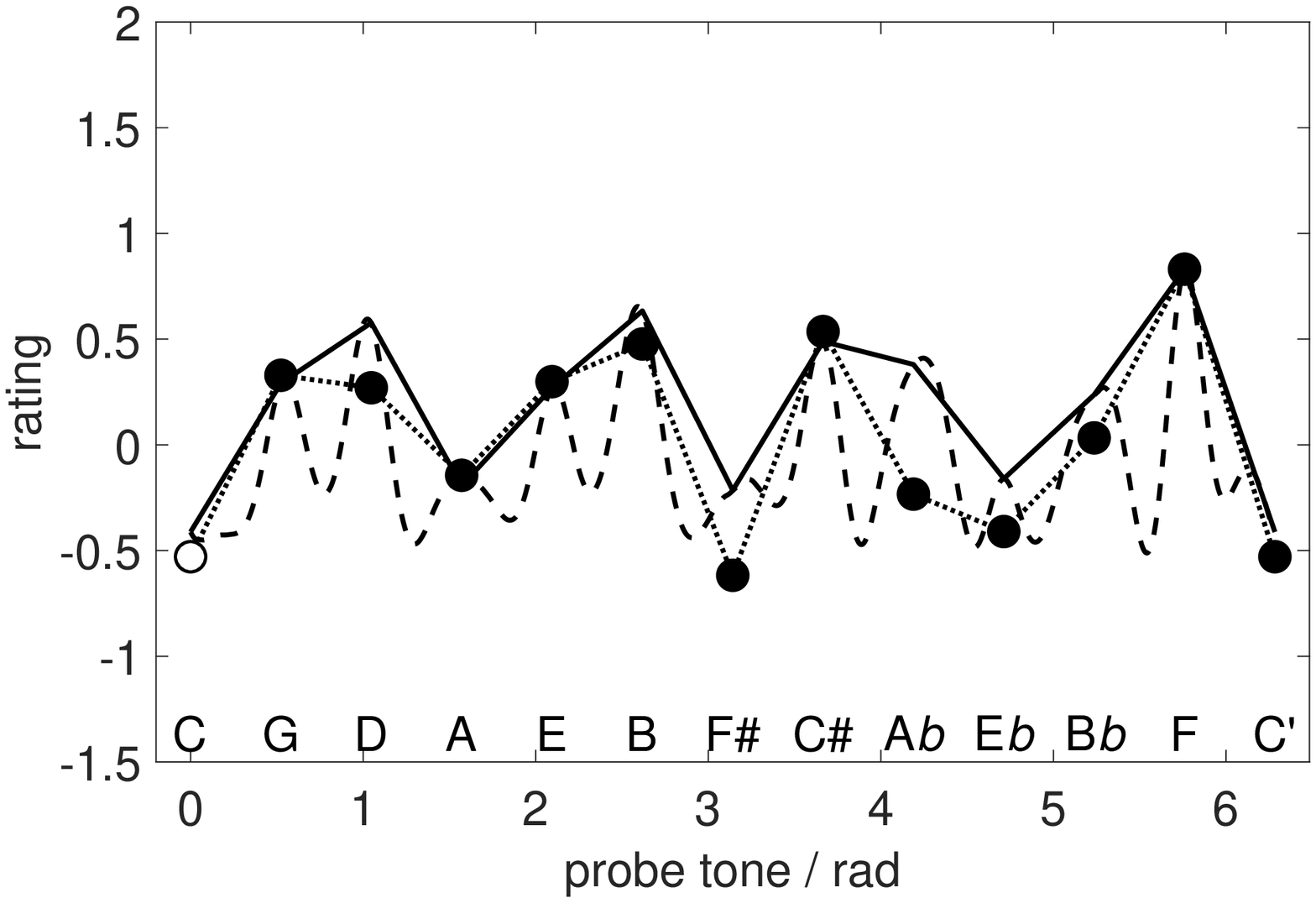}}
\subfigure[]{\includegraphics[scale=0.3]{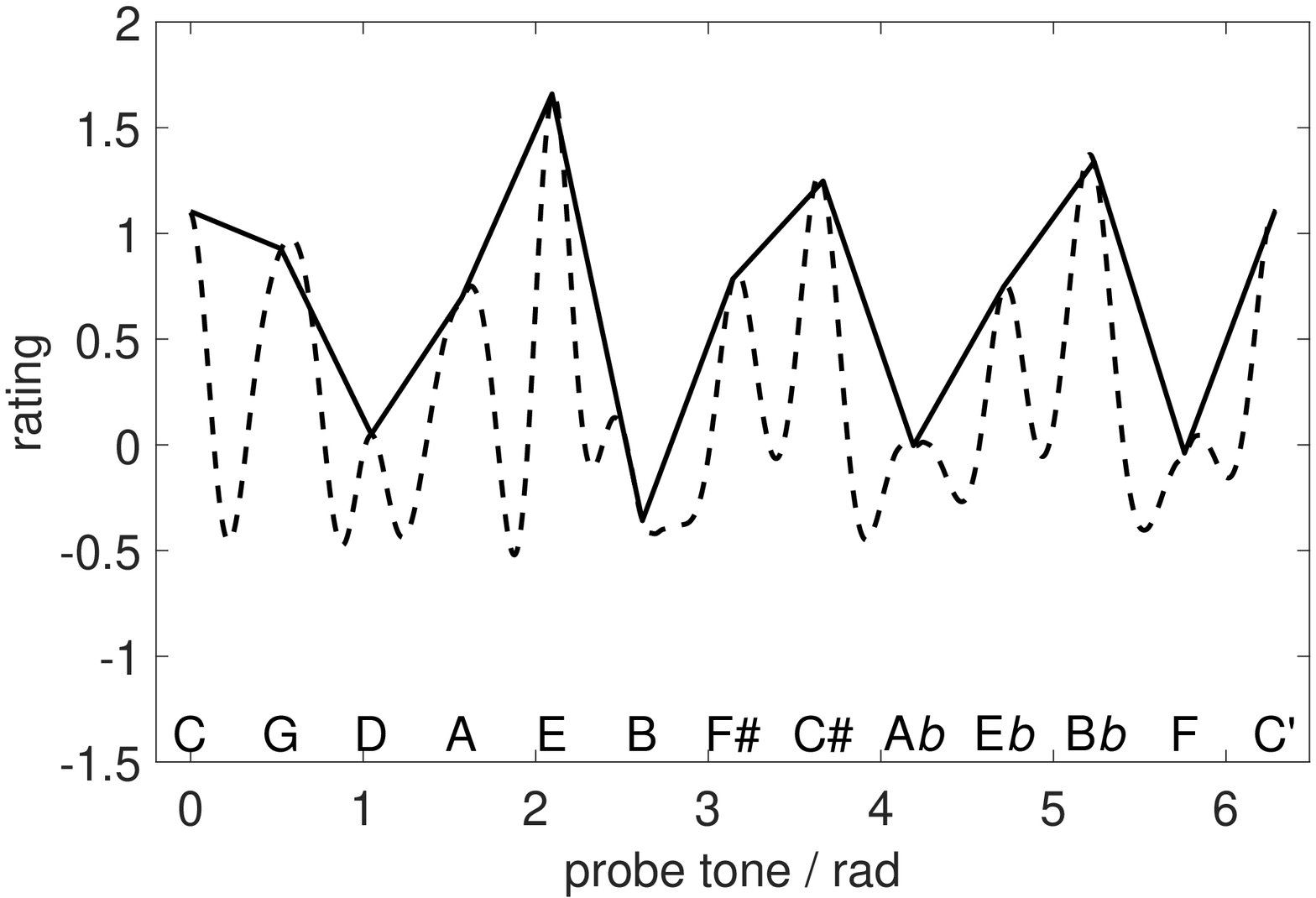}}
  \caption{\label{fig:W09mod} Quantum deformation model for dynamic tonal attraction data. Dotted with bullets: Rating data $A_\mathrm{W}(x)$ of \citet{Woolhouse09} against radian angles at the circle of fifths. Dashed: quantum probability density $p_C(x)$ of the mixed state model \pref{eq:wool3}. (a) For C major context. (b) For  C minor context. (c) For dominant seventh. (d) For French sixth. (e) For half-diminished seventh. (f) For ``Tristan chord''. Solid: mixed state density after sampling along the circle of fifths.}
\end{figure}

Here, we present the results of modeling the \citet{Woolhouse09} data with the quantum deformation model that was somewhat motivated by the interval cycle model. Computing the probability densities \pref{eq:wool3} for five contexts gives the results in \Fig{fig:W09mod}. Figure \ref{fig:W09mod}(a) shows the agreement of the model with the \citet{Woolhouse09} data for C major, (b) for C minor, (c) for the dominant seventh, (d) for French sixth, and (e) for half-diminished seventh. The dotted curves present the original results of the \citet{Woolhouse09} experiment. The dashed curves display the continuous quantum probability density $p_C$ obtained from the mixture of pure states over all context tones in the respective chords: major triad CEG (a), minor triad CE$\flat$G (b), dominant seventh CEGB$\flat$ (c), French sixth CEG$\flat$B$\flat$ (d), and half-diminished seventh CE$\flat$G$\flat$B$\flat$ (e).

Finally, we also deliver our model's prediction (f) for the famous ``Tristan chord'' from Richard Wagner's opera \emph{Tristan und Isolde} (1860). Following \citet{Woolhouse12}, the tension induced by the ``Tristan chord'' E$\flat$FG$\sharp$B, which is further strengthened by an intervening French sixth E$\flat$FAB, betokened by a single A, is eventually resolved by the chord EG$\sharp$DB$\flat$.\footnote{
     At the moment, no empirical data are available for the ``Tristan chord''.
}

The mixed state quantum deformation model assigns highest attraction values to F, A$\flat$, B, A, and C (in this order) [\Fig{fig:W09mod}(a)]. This is consistent with the interpretation of the C major context as the dominant of F major, containing F, A and C. Another possibility is the interpretation of the C major context as the subdominant of G major, comprised by G, B, and D. However, the agreement with model and data is confined to F, G, B$\flat$, and C$\sharp$. For C minor shown in [\Fig{fig:W09mod}(b)], the model almost perfectly agrees with the experimental results: highest attraction values are assigned to A$\flat$, F, and D. The model conforms also with music-theoretic predictions, resolving either the dominant of F minor into its tonic FA$\flat$C, or the subdominant of G minor into GB$\flat$D. Also for the dominant seventh [\Fig{fig:W09mod}(c)] the model correctly predicts F as the most likely resolution. For E and C$\sharp$ model and data agree quite well. The French sixth context [\Fig{fig:W09mod}(d)] correctly predicts resolution at B and F, but there is not much correlation with other data points. This is different for the half-diminished seventh plotted in [\Fig{fig:W09mod}(e)], where the model curve tracks the experimental data quite well, besides A$\flat$. Figure \ref{fig:W09mod}(f) demonstrates a remarkable performance of our model predicting the resolution of the ambiguous ``Tristan chord''. The predicted attraction rate is maximal at E and B$\flat$, those tones that belong to the resolving chord EG$\sharp$DB$\flat$. Interestingly, a similar analysis (not shown) for the intervening French sixth also yields large attraction values for E and B$\flat$, which are thereby further enhanced. In addition, the French sixth also increases attraction rates for the resolving G$\sharp$ and D. The ambiguity of the ``Tristan chord'' is reflected in our model by also predicting high attractions for C and C$\sharp$.\footnote{
    We thank one reviewer for this interesting remark.
}

The quality of our model is assessed by the regression analysis presented in \Tab{tab:dycor}.

\begin{table}[H]
  \centering
  \begin{tabular}{l*{2}{c}} \hline
  {}& IC model & deformation model \\
  C major & 0.69 & 0.44 \\
  C minor & 0.76 & 0.93 \\
  dominant seventh & 0.76 & 0.65 \\
  French sixth & 0.79 & 0.76 \\
  half-diminished seventh & 0.89 & 0.90 \\
  \hline
  \end{tabular}
    \caption{\label{tab:dycor} Correlation coefficients $r$ of two discussed models for dynamic tonal attraction with \citet{Woolhouse09} data. All error probabilities are significantly below 0.05.}
\end{table}

The correlation coefficients confirm the outcome of visually inspecting \Fig{fig:W09mod}. For C minor and the half-diminished seventh our quantum model performs slightly better than the IC model of \citet{Woolhouse09} and \citet{WoolhouseCross10}. Moreover, our model predicts the resolution of the ``Tristan chord'' \citep{Woolhouse12}. However, the main advantage of the quantum deformation model goes beyond the Woolhouse model and is able describing both the static and the dynamic attraction data.

\section{Discussion}
\label{sec:disc}

Tonal attraction is an important issue in the psychology of music. In this study, we have discussed two kinds of tonal attraction as investigated by probe tone experiments. Static tonal attraction refers to the stability or instability of tones or chords in a certain context, establishing a tonal key. By contrast, dynamic tonal attraction reflects the predictability of tones or chords continuing a preestablished melodic or harmonic context \citep{Temperley08}. In the paradigm of probe tone experiments \citep{KrumhanslShepard79, Krumhansl79}, static and dynamic attraction are investigated by means of different kinds of instructions: One the one hand, in a static attraction experiment, subjects are asked to rate how well a presented probe tone fits to a priming context \citep{KrumhanslKessler82}. In a dynamic attraction experiment, on the other hand, subjects are asked to rate how well a given probe tone completes or resolves the priming context, both melodically or harmonically \citep{Woolhouse09}.

It was the aim of this study to integrate structural and probabilistic theories of computational music theory into a unified framework. On the one hand, structural accounts such as the generative theory of tonal music are guided by principle musicological insights about the intrinsic symmetries of Western tonal music \citep{LerdahlJackendoff83, Lerdahl88, Lerdahl96}. On the other hand, probabilistic accounts such as Bayesian models or Gaussian Markov chains are able to describe melodic progession through statistical correlations in large music corpora \citep{Temperley07, Temperley08}. Here, we argued that quantum approaches to music cognition are able to unify symmetry and (quantum) probability in a single framework for data from static and dynamic attraction experiments.

In a first attempt to describing the static attraction data of \citet{KrumhanslKessler82}, we realized that simply rearranging the data points according to the universal circle of fifths, instead of increasing physical pitch frequency, revealed a systematic pattern of periodic attraction: tones close to the tonic are more attractive than tones in the vicinity of the tritone. Since this periodicity could be expressed as a cosine similarity measure, often used in quantum cognition models of similarity judgements \citep{PothosBusemayerTrueblood13, PothosTrueblood15}, we were led to the formulation of a quantum model of tonal attraction by means of a ``wave function'' defined over the circle group as ``configuration space'' which includes the circle of fifths as a subgroup. An important insight from our free quantum model is that its Hilbert space is essentially two-dimensional which proves its equivalence with an earlier qubit model proposed by \citet{Blutner15}.

In order to improve our free model, we developed a ``deformation'' quantum model, still based on cosine quantum similarity, but introducing a deformation of interval lengths over the circle of fifths. We described this deformation function as a symmetric polynomial of fourth order, and determined its parameters from two necessary interpolation conditions, saying that the tonic should not be deformed at all, while the tritone receives maximal deformation. The quantum wave functions of our approach can be interpreted as musical Gestalts in the sense of Gestalt psychology \citep{Kohler69}.

For assessment of the tonal attraction of complex priming stimuli such as chords or cadences, we followed a suggestion recently conjectured by \citet{WoolhouseCross10} and exploited by \citet{Blutner15} to compute the discrete convolution of a kernel function, the quantum probability obtained from the squared wave function, and a uniform distribution over all context tones. As contexts, we assumed the tonic triads of the C major and C minor keys for static attraction, here. This is similar to the hierarchical model of \citet{Lerdahl88, Lerdahl96}, where the tonic triad comprises level C in the hierarchy. As a result, our model precisely reproduced the experimental data with comparable statistical performance as the hierarchical model. However, while the hierarchical model has to stipulate the existence of all five levels in the hierarchy, our model only refers to the C level. The other levels A -- E required by the hierarchical model are deduced from our model. Therefore, our quantum model parsimoniously outperforms the hierarchical model of static attraction with respect to explanatory power.

However, the symmetric deformation model for static tonal attraction does not correctly describe the observed asymmetry between major and minor modes in the \citet{KrumhanslKessler82} data. This is also a problem for the hierarchical model as well. We therefore additionally presented an asymmetric deformation model that breaks the mirror symmetry against the tritone. Fitting two free parameters of this model to the major context, we were able to achieve substantial improvement also for minor keys and an understanding for the different degrees of consonance for major, minor, diminished, and augmented chords.

For dynamic tonal attraction we devised a similar quantum model based on a sixth-order polynomial deformation. This construction was guided by the interval cycle model of \citet{WoolhouseCross10} and \citet{Woolhouse09} and in agreement to related models that prefer small intervals over longer ones for melodic and harmonic dynamics \citep{CurtisBharucha09, KrumhanslEA00, Narmour92, Temperley08}. Our model performed comparably well as the ICP model in a regression analysis. Interestingly, our model confirms predictions about the resolution of context chords based on musical harmony theory with good accuracy.

Finally, we address the possible relevance of our work to the cognitive neuroscience of music perception. The behavioral findings of \citet{KrumhanslKessler82} and \citet{Woolhouse09} have been supported by neuropsychological experiments in the event-related brain potential (ERP) \citep{GranotHai09, Limb06} and in the functional magnetic resonance (fMRI) paradigms \citep{DurrantMirandaEA07, JanataEA02, KoelschGunterEA02, Limb06, VaqueroHartmannEA16}. The first three fMRI studies used melodies with harmonic modulation from one key into another key as stimuli and reported significant anatomical differences for key changes. In particular, \citet{JanataEA02}, who carried out a regression analysis of fMRI data with self-organized maps trained upon the third torus as tonal space \citep{JanataEA02, PurwinsBlankertzObermayer07}, made the strong claim that the rostromedial prefrontal cortex exhibits a neural tonotopic representation of the third torus.

\section{Conclusions}
\label{sec:conc}

This study applies methods from the quantum cognition framework to tonal attraction. In probe tone experiments, music psychologists measure the likelihood of chromatic probe tones relative to a priming context. Depending on the particular instruction, the subject's ratings assess either the degree of fit between a probe tone and a key context (static attraction), or the degree of predictability of a probe tone given a preceding context (dynamic attraction). A first attempt reveals that tonal attraction correlates with quantum similarity between tones across the universal circle of fifths. Deforming the distances between tones leads to two quantum wave functions, one for static attraction, the second for dynamic attraction. We compute their attraction profiles for tonic contexts and compare them with predictions of the hierarchical model in the static case and of the interval cycle model for the dynamic case. Our parsimonious quantum models provide precise and rigorous results with great explanatory power.

\section*{Acknowledgments}

We thank Maria Mannone, Serafim Rodrigues and G\"unther Wirsching for valuable comments on an earlier version of the manuscript.

\section*{References}


\begin{thebibliography}{}

\bibitem[Balzano, 1980]{Balzano80}
Balzano, G.~J. (1980).
\newblock The group-theoretic description of 12-fold and microtonal pitch
  systems.
\newblock {\em Computer Music Journal}, 4(4):66 -- 84.

\bibitem[beim Graben and Blutner, 2017]{GrabenBlutner17}
beim Graben, P. and Blutner, R. (2017).
\newblock Toward a gauge theory of musical forces.
\newblock In de~Barros, J.~A., Coecke, B., and Pothos, E., editors, {\em
  Quantum Interaction. 10th International Conference (QI 2016)}, volume 10106
  of {\em LNCS}, pages 99 -- 111. Springer, Cham.

\bibitem[Blutner, 2015]{Blutner15}
Blutner, R. (2015).
\newblock Modelling tonal attraction: tonal hierarchies, interval cycles, and
  quantum probabilities.
\newblock {\em Soft Computing}, pages 1 -- 19.

\bibitem[Blutner and beim Graben, 2016]{BlutnerGraben16}
Blutner, R. and beim Graben, P. (2016).
\newblock Quantum cognition and bounded rationality.
\newblock {\em Synthese}, 193:3239 -- 3291.

\bibitem[Busemeyer and Bruza, 2012]{BusemeyerBruza12}
Busemeyer, J.~R. and Bruza, P.~D., editors (2012).
\newblock {\em Quantum Models of Cognition and Decision}.
\newblock Cambridge University Press.

\bibitem[Curtis and Bharucha, 2009]{CurtisBharucha09}
Curtis, M.~E. and Bharucha, J.~J. (2009).
\newblock Memory and musical expectation for tones in cultural context.
\newblock {\em Music Perception: An Interdisciplinary Journal}, 26(4):365 --
  375.

\bibitem[Durrant et~al., 2007]{DurrantMirandaEA07}
Durrant, S., Miranda, E.~R., Hardoon, D., Shawe-Taylor, J., Brechmann, A., and
  Scheich, H. (2007).
\newblock Neural correlates of tonality in music.
\newblock In {\em Proceedings of the NIPS Conference Workshop on Music, Brain
  \& Cognition}.

\bibitem[Granot and Hai, 2009]{GranotHai09}
Granot, R.~Y. and Hai, A. (2009).
\newblock Electrophysiological evidence for a two-stage process underlying
  single chord priming.
\newblock {\em NeuroReport}, 20(9).

\bibitem[Janata, 2007]{Janata07}
Janata, P. (2007).
\newblock Navigating tonal space.
\newblock In Selfridge-Field, E. and Hewlett, W.~B., editors, {\em Tonal Theory
  for the Digital Age}, volume~15 of {\em Computing in Musicology}, 
  pages 39 -- 50. MIT Press, Cambridge (MA).

\bibitem[Janata et~al., 2002]{JanataEA02}
Janata, P., Birk, J.~L., Horn, J. D.~V., Leman, M., Tillmann, B., and Bharucha,
  J.~J. (2002).
\newblock The cortical topography of tonal structures underlying western music.
\newblock {\em Science}, 298(5601):2167 -- 2170.

\bibitem[Johnson-Laird et~al., 2012]{JohnsonKangLeong12}
Johnson-Laird, P.~N., Kang, O.~E., and Leong, Y.~C. (2012).
\newblock On musical dissonance.
\newblock {\em Music Perception: An Interdisciplinary Journal}, 30(1):19 -- 35.

\bibitem[Koelsch et~al., 2002]{KoelschGunterEA02}
Koelsch, S., Gunter, T.~C., von Cramon, D.~Y., Zysset, S., Lohmann, G., and
  Friederici, A.~D. (2002).
\newblock {Bach speaks: a cortical ``language-network'' serves the processing
  of music}.
\newblock {\em NeuroImage}, 17(2):956 -- 966.

\bibitem[K\"ohler, 1969]{Kohler69}
K\"ohler, W. (1969).
\newblock {\em The Task of Gestalt Psychology}.
\newblock Princeton University Press, New Jersey.

\bibitem[Krumhansl, 1979]{Krumhansl79}
Krumhansl, C.~L. (1979).
\newblock The psychological representation of musical pitch in a tonal context.
\newblock {\em Cognitive Psychology}, 11(3):346 -- 374.

\bibitem[Krumhansl and Cuddy, 2010]{KrumhanslCuddy10}
Krumhansl, C.~L. and Cuddy, L.~L. (2010).
\newblock A theory of tonal hierarchies in music.
\newblock In Jones, M.~R., Fay, R.~R., and Popper, A.~N., editors, {\em Music
  Perception}, Springer Handbook of Auditory Research, pages 51 --
  87. Springer, New York.

\bibitem[Krumhansl and Kessler, 1982]{KrumhanslKessler82}
Krumhansl, C.~L. and Kessler, E.~J. (1982).
\newblock Tracing the dynamic changes in perceived tonal organization in a
  spatial representation of musical keys.
\newblock {\em Psychological Review}, 89(4):334 -- 368.

\bibitem[Krumhansl and Shepard, 1979]{KrumhanslShepard79}
Krumhansl, C.~L. and Shepard, R.~N. (1979).
\newblock Quantification of the hierarchy of tonal functions within a diatonic
  context.
\newblock {\em Journal of Experimental Psychology: Human Perception and
  Performance}, 5(4):579.

\bibitem[Krumhansl et~al., 2000]{KrumhanslEA00}
Krumhansl, C.~L., Toivanen, P., Eerola, T., Toiviainen, P., J\"{a}rvinen, T.,
  and Louhivuori, J. (2000).
\newblock {Cross-cultural music cognition: Cognitive methodology applied to
  North Sami yoiks}.
\newblock {\em Cognition}, 76(1):13 -- 58.

\bibitem[Lerdahl, 1988]{Lerdahl88}
Lerdahl, F. (1988).
\newblock Tonal pitch space.
\newblock {\em Music Perception: An Interdisciplinary Journal}, 5:315 -- 350.

\bibitem[Lerdahl, 1996]{Lerdahl96}
Lerdahl, F. (1996).
\newblock Calculating tonal tension.
\newblock {\em Music Perception: An Interdisciplinary Journal}, 13(3):319 --
  363.

\bibitem[Lerdahl, 2015]{Lerdahl15}
Lerdahl, F. (2015).
\newblock Concepts and representations of musical hierarchies.
\newblock {\em Music Perception: An Interdisciplinary Journal}, 33(1):83 -- 95.

\bibitem[Lerdahl and Jackendoff, 1983]{LerdahlJackendoff83}
Lerdahl, F. and Jackendoff, R. (1983).
\newblock {\em A Generative Theory of Tonal Music}.
\newblock MIT Press, Cambridge (MA).

\bibitem[Limb, 2006]{Limb06}
Limb, C.~J. (2006).
\newblock Structural and functional neural correlates of music perception.
\newblock {\em The Anatomical Record Part A: Discoveries in Molecular,
  Cellular, and Evolutionary Biology}, 288A(4):435 -- 446.

\bibitem[Mannone, 2018]{Mannone18a}
Mannone, M. (2018).
\newblock {Knots, music, and DNA}.
\newblock {\em Journal of Creative Music Systems}, 2(2).

\bibitem[Mannone and Compagno, 2013]{MannoneCompagno13}
Mannone, M. and Compagno, G. (2013).
\newblock Characterization of the degree of musical {non-Markovianity}.
\newblock arXiv:1306.0229 [physics.data-an].

\bibitem[Mannone and Mazzola, 2015]{MannoneMazzola15}
Mannone, M. and Mazzola, G. (2015).
\newblock Hypergestures in complex time: Creative performance between symbolic
  and physical reality.
\newblock In Collins, T., Meredith, D., and Volk, A., editors, {\em Proceedings
  of the 5th International Conference on Mathematics and Computation in Music
  (MCM 2015)}, pages 137 -- 148, Cham. Springer.

\bibitem[Mazzola, 1990]{Mazzola90}
Mazzola, G. (1990).
\newblock {\em Geometrie der T\"{o}ne}.
\newblock Birkh\"{a}user, Basel.

\bibitem[Mazzola, 2002]{Mazzola02}
Mazzola, G. (2002).
\newblock {\em The Topos of Music: Geometric Logic of Concepts, Theory, and
  Performance}.
\newblock Birkh{\"a}user, Basel.

\bibitem[Mazzola et~al., 2016]{MazzolaMannonePang16}
Mazzola, G., Mannone, M., and Pang, Y. (2016).
\newblock {\em Cool Math for Hot Music}.
\newblock Computational Music Science. Springer, Cham.

\bibitem[Milne et~al., 2015]{MilneLaneySharp15}
Milne, A.~J., Laney, R., and Sharp, D.~B. (2015).
\newblock A spectral pitch class model of the probe tone data and scalic
  tonality.
\newblock {\em Music Perception: An Interdisciplinary Journal}, 32(4):364 --
  393.

\bibitem[Milne et~al., 2011]{MilneEA11}
Milne, A.~J., Sethares, W.~A., Laney, R., and Sharp, D.~B. (2011).
\newblock Modelling the similarity of pitch collections with expectation
  tensors.
\newblock {\em Journal of Mathematics and Music}, 5(1):1 -- 20.

\bibitem[Narmour, 1992]{Narmour92}
Narmour, E. (1992).
\newblock {\em The Analysis and Cognition of Melodic Complexity: The
  Implication-Realization Model}.
\newblock University of Chicago Press, Chicago (IL).

\bibitem[Parncutt, 1989]{Parncutt89}
Parncutt, R. (1989).
\newblock {\em Harmony: A Psychoacoustical Approach}.
\newblock Springer, Berlin.

\bibitem[Parncutt, 2011]{Parncutt11}
Parncutt, R. (2011).
\newblock The tonic as triad: Key profiles as pitch salience profiles of tonic
  triads.
\newblock {\em Music Perception: An Interdisciplinary Journal}, 28(4):333 --
  366.

\bibitem[Pothos and Busemeyer, 2013]{PothosBusemeyer13}
Pothos, E.~M. and Busemeyer, J.~R. (2013).
\newblock Can quantum probability provide a new direction for cognitive
  modeling?
\newblock {\em Behavioral and Brain Sciences}, 36:255 -- 274.

\bibitem[Pothos et~al., 2013]{PothosBusemayerTrueblood13}
Pothos, E.~M., Busemeyer, J.~R., and Trueblood, J.~S. (2013).
\newblock A quantum geometric model of similarity.
\newblock {\em Psychological Review}, 120(3):679 -- 696.

\bibitem[Pothos and Trueblood, 2015]{PothosTrueblood15}
Pothos, E.~M. and Trueblood, J.~S. (2015).
\newblock Structured representations in a quantum probability model of
  similarity.
\newblock {\em Journal of Mathematical Psychology}, 64-65:35 -- 43.

\bibitem[Purwins et~al., 2007]{PurwinsBlankertzObermayer07}
Purwins, H., Blankertz, B., and Obermayer, K. (2007).
\newblock Toroidal models in tonal theory and pitch-class analysis.
\newblock In Selfridge-Field, E. and Hewlett, W.~B., editors, {\em Tonal Theory
  for the Digital Age}, volume~15 of {\em Computing in Musicology}, 
  pages 73 -- 98. MIT Press, Cambridge (MA).

\bibitem[Quinn, 2010]{Quinn10}
Quinn, I. (2010).
\newblock {On Woolhouse's interval-cycle proximity hypothesis}.
\newblock {\em Music Theory Spectrum}, 32:172 -- 179.

\bibitem[Sch\"{o}nberg, 1978]{Schonberg78}
Sch\"{o}nberg, A. (1978).
\newblock {\em Theory of Harmony}.
\newblock University of California Press, Berkeley (CA).

\bibitem[Stolzenburg, 2015]{Stolzenburg15}
Stolzenburg, F. (2015).
\newblock Harmony perception by periodicity detection.
\newblock {\em Journal of Mathematics and Music}, 9(3):215 -- 238.

\bibitem[Temperley, 2007]{Temperley07}
Temperley, D. (2007).
\newblock {\em Music and Probability}.
\newblock MIT Press, Cambridge (MA).

\bibitem[Temperley, 2008]{Temperley08}
Temperley, D. (2008).
\newblock A probabilistic model of melody perception.
\newblock {\em Cognitive Science}, 32(2):418 -- 444.

\bibitem[v.~Helmholtz, 1877]{Helmholtz1877}
v.~Helmholtz, H. (1877).
\newblock {\em On the Sensations of Tones}.
\newblock Dover, New York (NY).
\newblock Translated by A. J. Ellis.

\bibitem[Vaquero et~al., 2016]{VaqueroHartmannEA16}
Vaquero, L., Hartmann, K., Ripoll\'{e}s, P., Rojo, N., Sierpowska, J.,
  Fran\c{c}ois, C., C\`{a}mara, E., van Vugt, F.~T., Mohammadi, B., Samii, A.,
  M\"{u}nte, T.~F., Rodr\'{i}guez-Fornells, A., and Altenm\"{u}ller, E. (2016).
\newblock Structural neuroplasticity in expert pianists depends on the age of
  musical training onset.
\newblock {\em NeuroImage}, 126(Supplement C):106 -- 119.

\bibitem[Woolhouse, 2009]{Woolhouse09}
Woolhouse, M. (2009).
\newblock Modelling tonal attraction between adjacent musical elements.
\newblock {\em Journal of New Music Research}, 38(4):357 -- 379.

\bibitem[Woolhouse, 2012]{Woolhouse12}
Woolhouse, M. (2012).
\newblock Wagner in the round: using interval cycles to model chromatic
  harmony.
\newblock In Cambouropoulos, E., Tsougras, C., Mavromatis, P., and Pastiadis,
  K., editors, {\em Proceedings of the 12th International Conference on Music
  Perception and Cognition and the 8th Triennial Conference of the European
  Society for the Cognitive Sciences of Music}, pages 1142 -- 1145.

\bibitem[Woolhouse and Cross, 2010]{WoolhouseCross10}
Woolhouse, M. and Cross, I. (2010).
\newblock {Using interval cycles to model Krumhansl's tonal hierarchies}.
\newblock {\em Music Theory Spectrum}, 32(1):60 -- 78.

\end{thebibliography}

\end{document}